\documentclass[10pt,journal,compsoc]{IEEEtran}
\usepackage[T1]{fontenc}
\usepackage[utf8]{inputenc}
\usepackage[table,dvipsnames]{xcolor}
\usepackage{CJKutf8}
\usepackage{graphicx}
\usepackage{xspace}
\usepackage{tabularx}
\usepackage{pgfplots}
\usepackage{pgfplotstable}
\pgfplotsset{compat=1.14} % to remove warning

\usepackage{tikz}
\def\checkmark{\tikz\fill[scale=0.4](0,.35) -- (.25,0) -- (1,.7) -- (.25,.15) -- cycle;}

\usepackage{listings}
\usepackage{booktabs}
\usepackage{rotating}
\usepackage[inline]{enumitem}
\usepackage[autolanguage]{numprint}
\usepackage{mdframed}

\definecolor{diffstart}{RGB}{159,177,186}
\definecolor{diffincl}{RGB}{5,205,107}
\definecolor{diffrem}{RGB}{255,33,93}

\definecolor{bluekeywords}{rgb}{0.13,0.13,1}
\definecolor{greencomments}{rgb}{0,0.5,0}
\definecolor{redstrings}{rgb}{0.9,0,0}
\definecolor{mygray}{rgb}{0.5,0.5,0.5}

\newcommand{\lstbg}[3][0pt]{{\fboxsep#1\colorbox{#2}{\strut #3}}}

\lstdefinelanguage{XML}
{
  morestring=[b]",
  morestring=[s]{>}{<},
  morecomment=[s]{<?}{?>},
  stringstyle=\color{black},
  identifierstyle=\color{bluekeywords},
  keywordstyle=\color{cyan},
  morekeywords={xmlns,version,type}% list your attributes here
}

\lstdefinelanguage{diff}{
morecomment=[f][\lstbg{diffstart}]{@@},
morecomment=[f][\lstbg{green!20}]{+\ },
morecomment=[f][\lstbg{red!20}]{-\ },
}

\mdfdefinestyle{mpdframe}{
    frametitlebackgroundcolor   =black!15,
    frametitlerule              =true,
    roundcorner                 =5pt,
    middlelinewidth             =0.8pt,
    innermargin                 =0.2cm,
    outermargin                 =0.2cm,
    innerleftmargin             =0.2cm,
    innerrightmargin            =0.2cm,
    innertopmargin              =0.2cm,
    innerbottommargin           =0.2cm
}

\newcommand{\argo}{\textsc{Argo}\xspace}
\newcommand{\placebo}{\textsc{Placebo}\xspace}
\newcommand{\json}{JSON\xspace}

\usepackage{epigraph}

\usepackage{fp}
\usepackage{xspace}
\usepackage{numprint}
\newcommand*\np[2][z]{%\textcolor{red}{%
\ifx z#1%
$\numprint{#2}$%
\else%
$\numprint[#1]{#2}$%
\fi\xspace%}
}

\newcommand{\ShowAbsoluteNumber}[1]{%
\ifnum #1<10%
{\hspace*{0pt}#1}%
\else%
\ifnum #1<100%
{\hspace*{0pt}#1}%
\else%
\ifnum #1<1000%
{\hspace*{0pt}#1}%
\else%
{\numprint{#1}}%
\fi%
\fi%
\fi%
}

\newcommand{\ShowPercentage}[2]{%
\FPeval\percentage{round(#1/#2*100,0)}%
\FPeval\percentageOneDecimal{round(#1/#2*100,1)}%
\ifnum \percentage=0%
{\np[\%]{\FPprint{percentageOneDecimal}}}%
\else%
\ifnum \percentage<10%
{\np[\%]{\FPprint{percentageOneDecimal}}}%
\else%
{\np[\%]{\FPprint{percentageOneDecimal}}}%
\fi%
\fi%
\xspace
}
\newlength\BARSIZE  
\newcommand{\inlinechart}[2]{%
\FPeval{\BLACKBARSIZE}{#1/#2}\textcolor{black!80}{\rule{\BLACKBARSIZE\BARSIZE}{1.6ex}}%
\FPeval{\BLACKBARSIZE}{1 - (#1/#2)}\textcolor{black!10}{\rule{\BLACKBARSIZE\BARSIZE}{1.6ex}}%
}

\newcommand*\ChartSmall[3][v]{%
\ifx q#1%
    \np{#2}/\np{#3}(\ShowPercentage{#2}{#3})\else%
\ifx p#1%
    \np{#2}(\ShowPercentage{#2}{#3})\else%
\ifx c#1%
    \inlinechart{#2}{#3}%
\else%
    \np{#2}%
    \ifx r#1%
        /\np{#3}%
    \fi%
    \hspace*{0.5ex}(\ShowPercentage{#2}{#3}) %
    \inlinechart{#2}{#3}%
    \xspace
\fi\fi\fi%
}

\usepackage[]{hyperref}

\begin{document}

\author{
\IEEEauthorblockN{Nicolas Harrand\IEEEauthorrefmark{1}, Thomas Durieux\IEEEauthorrefmark{1}, David Broman\IEEEauthorrefmark{1}\IEEEauthorrefmark{2}, and Benoit Baudry\IEEEauthorrefmark{1}}

\IEEEauthorblockA{
\IEEEauthorrefmark{1}\textit{EECS, KTH Royal Institute of Technology, Stockholm, Sweden}\\ Email: \{harrand, tdurieux, dbro, baudry\}@kth.se\\
}
\IEEEauthorrefmark{2}\textit{
Digital Futures
}
}

%\title{Automatic Diversification of Java Applications through Library Replacement: the case of JSON Libraries}
%\title{\argo: Automatic diversification of \json Software Suppliers}
\title{Automatic Diversity in the Software Supply Chain}
%\title{Harnessing the Diversity of JSON Libraries in the Software Supply Chain}
%\argo: Harnessing the Natural of Diversity of JSON libraries to automatically diversify Java applications}
%\title{Argo: Automatic diversification of JSON libraries in Java applications }
%Contrarily to the ship of Theseus, the ship of JaSON is definitly not the same when it comes back to the harbor.

\IEEEtitleabstractindextext{
\begin{abstract}
Despite its obvious benefits, the increased adoption of package managers to automate the reuse of libraries has opened the door to a new class of hazards: supply chain attacks. By injecting malicious code in one library, an attacker may compromise all instances of all applications that depend  on the library. To mitigate the impact of supply chain attacks, we propose the concept of Library Substitution Framework. This novel concept leverages one key observation: when an application depends on a library, it is very likely that there exists other libraries that provide similar features. The key objective of Library Substitution Framework is to enable the developers of an application to harness this diversity of libraries in their supply chain. The framework lets them generate a population of application variants, each depending on a different alternative library that provides similar functionalities. To investigate the relevance of this concept, we develop \argo, a proof-of-concept implementation of this framework that harnesses the diversity of \json suppliers. We study the feasibility of library substitution and its impact on a set of \nballclient clients. Our empirical results show that for \nbmassdivclient of the \nballclient java applications tested, we can substitute the original \json library used by the client by at least $15$ other \json libraries without modifying the client's code. These results show the capacity of a Library Substitution Framework to diversify the supply chain of the client applications of the libraries it targets.
\end{abstract}
\begin{IEEEkeywords}
Software supply chain, Library Substitution, Software repository, Software reuse, Java, Maven Central Repository
\end{IEEEkeywords}}

\IEEEdisplaynontitleabstractindextext
\IEEEpeerreviewmaketitle

\colorlet{punct}{red!60!black}
\definecolor{background}{HTML}{EEEEEE}
\definecolor{delim}{RGB}{20,105,176}
\colorlet{numb}{magenta!60!black}

\lstdefinelanguage{json}{
    basicstyle=\normalfont\ttfamily,
    numbers=left,
    numberstyle=\scriptsize,
    stepnumber=1,
    numbersep=8pt,
    showstringspaces=false,
    breaklines=true,
    literate=
     *{0}{{{\color{numb}0}}}{1}
      {1}{{{\color{numb}1}}}{1}
      {2}{{{\color{numb}2}}}{1}
      {3}{{{\color{numb}3}}}{1}
      {4}{{{\color{numb}4}}}{1}
      {5}{{{\color{numb}5}}}{1}
      {6}{{{\color{numb}6}}}{1}
      {7}{{{\color{numb}7}}}{1}
      {8}{{{\color{numb}8}}}{1}
      {9}{{{\color{numb}9}}}{1}
      {:}{{{\color{punct}{:}}}}{1}
      {,}{{{\color{punct}{,}}}}{1}
      {\{}{{{\color{delim}{\{}}}}{1}
      {\}}{{{\color{delim}{\}}}}}{1}
      {[}{{{\color{delim}{[}}}}{1}
      {]}{{{\color{delim}{]}}}}{1},
}

%Dataset
\def\nbTotalProject{147991}

%Numbers
\newcommand{\nbsrcjson}{$4$\@\xspace}
\newcommand{\nbtargetjson}{$20$\@\xspace}

\newcommand{\nbcorrectjson}{$211$\@\xspace}
\newcommand{\nberroredjson}{$281$\@\xspace}
\newcommand{\nballjson}{$492$\@\xspace}

\newcommand{\nballclient}{$368$\@\xspace}
\newcommand{\nbapicompclient}{$329$\@\xspace}
\newcommand{\nbdivclient}{$272$\@\xspace}
\newcommand{\nbmassdivclient}{$195$\@\xspace}

%Names
\newcommand{\orgjson}{\texttt{org.json}\@\xspace}
\newcommand{\jsonsimple}{\texttt{json-simple}\@\xspace}
\newcommand{\gson}{\texttt{gson}\@\xspace}
\newcommand{\fastjson}{\texttt{fastjson}\@\xspace}
\newcommand{\jackson}{\texttt{jackson-databind}\@\xspace}

\newcommand{\wfcorpus}{\texttt{Well-formed}\@\xspace}
\newcommand{\ifcorpus}{\texttt{Ill-formed}\@\xspace}

\newcommand{\yasjf}{\texttt{argo}\@\xspace}
\newcommand{\yasjfapi}{\texttt{argo-api}\@\xspace}

\maketitle

\section{Introduction}
%Introduce software supply chains with emphasis on online repository of dependencies.
Modern software development increasingly relies on the reuse of external libraries. The past decades have seen the emergence of online software repositories hosting thousands, sometimes millions, of software artifacts ready for reuse~\cite{decan19}. Package managers that automate library resolution from these repositories have greatly facilitated the practice~\cite{cox2019surviving}. Increasingly, modern software consists, in a majority of external libraries glued together by a minority of business specific code directly written by the developer~\cite{Holger}. The benefit of relying on externally developed software artifacts is twofold: development and maintenance time is decreased, and software used by thousands of other applications may receive more scrutiny. %For instance, it is technically possible to rewrite a JSON parser from scratch for an application that wishes to read a simple configuration file written in this format. Yet, the cost associated with developing, testing and maintaining a parser, compared to simply reusing one of the many existing ones, makes it an option unlikely to be selected by developers. 

%Introduce supply chain hazard
However, depending on third-party libraries opens up for a whole new category of hazards.  Bugs~\cite{carvalho2014heartbleed}, breakages~\cite{leftpad}, and supply chain attacks~\cite{ohm20,taylor2020defending} can now occur in all the dependencies of an application and not solely in the code controlled by the application developer. 
%Supply chain hazard examples
For example, in March 2016, the removal of the \texttt{left-pad} package from the NPM registry broke the build of thousands of JavaScript projects ~\cite{leftpad,cox2019surviving}. 
More recently, thousands of SolarWinds customers have been compromised through a malicious update. This incident raised questions on how to reliably use software dependencies~\cite{MassacciJP21}.

The breadth of supply chain attacks is a major challenge. Compromising one single library ~\cite{eventstream} affects every instance of any application depending on the library. This makes popular libraries target of choices for malicious actors. On the other hand, seminal work by Forrest~\cite{forrest97} and Cohen~\cite{cohen93} have proposed to distribute software variants instead of running identical instances across a network of machine, to mitigate the scale of attacks.

We propose to diversify the libraries in the supply chain of applications to mitigate the scale of supply chain attacks.
For many application domains, there exist several alternative libraries providing similar features. This opens the door to the generation of variants of an application by substituting libraries in its supply chain.
Unfortunately, these libraries cannot be directly substituted.  They typically do not provide the same API ~\cite{bartolomei2009study}: features might not be divided in the same way, signatures may be different, etc. Hence, replacing one by another, requires considerable effort. %Approaches based on modifying clients' sources have been tried before~\cite{alrubayeM019,balaban2005refactoring}, but none of them are fully automated. This difficulty of migration makes clients tied to their libraries.

In this work, we introduce the concept of Library Substitution Framework. To support the substitution of libraries that provide similar features with different APIs, without modifying the application code, we propose to separate the API from the actual implementation. Keeping the API intact is necessary to prevent changes in the application. To bind the API to an actual implementation, we propose a three-tier adaptation architecture. First, a bridge provides the same API as the original and maps this API to an abstract Facade. The Facade captures the core features that are common to the pool of similar libraries. The third-tier for adaptation relates the Facade to an actual implementation. For a specific library, a wrapper implements this relation to the Facade's features. With this three-tier architecture we save the effort of developing one adapter per combination of possible library substitution.

To assess the feasibility of this architecture, diversifying supply chains, we implement a concrete framework for Java \json libraries. The framework, called \argo, supports substituting a \json library present in the supply chain of an application, by another \json library. 
We discuss the implementation process of \argo, its design choices, and experiment the reuse of the test suites of existing \json libraries to validate it.
%Through this proof of concept, we explore the consequences of trade\-offs between the share of API adapted and the share of clients that can be transformed. 
We also assess the effect of library substitution on 368 clients of \json libraries. We successfully generate  \np{4069} variant applications that use a completely different implementation of a \json library than the original and still pass the same tests as the original. For all these case studies, we have checked that the tests of the applications invoke a part of the \json API. These novel results demonstrate the opportunity of harnessing the natural diversity of library implementations in a Library Substitution Framework to diversify the supply chain of applications. Our library substitution architecture is generic and can be reused to diversify other parts of the supply chain for which there exist diverse implementations.

Our contributions are as follows
\begin{itemize}
    \item an original architecture for a Library Substitution Framework to diversify the supply chain of an application, without changing the application's source code;
    \item \argo, a proof-of-concept implementation of this framework to provide build time diversification of \json suppliers;
    \item novel empirical evidence that a Library Substitution Framework can generate application variants that have different dependencies as the original applications and still behave the same, modulo test suite.
\end{itemize}

%The remaining of this paper is organized as follows: \autoref{sec:background} presents background elements, \autoref{sec:concept} presents the goals, challenges, and the architecture of a Library Substitution Framework; \autoref{sec:methodology} details the methodology followed to evaluate this architecture and its impact on clients; \autoref{sec:results} contains the empirical results of this evaluation; \autoref{sec:related-work} presents related work; and \autoref{sec:conclusion} concludes.

\section{Risks of monoculture in the software supply chain}
\label{sec:background}

%\david{Intro is missing. The title does not directly match the overview paragraph in the end of section 1.}
%In this section we discuss how package managers and their associated repository have given birth to birth to a
This section introduces the context of our work. We summarize the challenges of library reuse, with respect to software supply chain attacks.

\subsection{Software Supply Chain Attacks}

Package managers  automatically fetch  third-party libraries from a  repository and integrate them into applications~\cite{decan19,cox2019surviving}. 
There exist package managers and an associated repository for most programming languages. For instance, Java has Maven and Maven Central, JavaScript has NPM and the NPM repository, Python has pip and PyPI. 

Massive dependence on external software artifacts raises new reliability and security concerns. New external actors become involved in the development of software projects: the developers of all external libraries used in the project. 
Furthermore, external libraries come with dependencies of their own, making the complete audit of libraries more difficult. Consequently, third-party software libraries represent an essential and complex part of the software supply chain of applications~\cite{ohm20,levy2003poisoning}.
This opens the door to software supply chain attacks, that specifically target third-party packages \cite{ohm20}.

Such attacks include package typo squatting \cite{taylor2020defending}, trusting the trust attacks \cite{thompson1984reflections}, and social engineering on open-source packages.
An example of such an attack is demonstrated by the incident related to the \texttt{event-stream} package from the NPM registry. A malicious actor managed to gain the trust of the owner by contributing legitimate patches to the package~\cite{eventstream}. The owner then handed over the control of the repository to the malicious actor, who used it to inject code stealing cryptocurrency keys available on the machine of anyone running code depending on the \texttt{event-stream} package.

%Hence, popular packages are excellent targets for supply chain attacks from malicious actors wishing to have a wide impact.

\subsection{Diversity Reservoirs in Software Repositories}
\label{sec:back-json-libs}
Software repositories host millions of libraries. Among these numerous libraries, many of them provide similar features.
For example, \textit{\url{mvnrepository.com}} lists more than $80$ different HTTP clients libraries for the JVM, more than $40$ different logging frameworks, and more than $20$ Base64 libraries. These different libraries that provide similar functionalities, are developed by different developers, with different motivations and without coordination. In the remaining of this paper, we call such a group of libraries implementing similar features, a \textit{reservoir}.

The libraries in a reservoir do not provide the exact same features, but very similar ones,  typically related to a standard. 
For example, network libraries  implement well-specified protocols (HTTP, TCP, IP), cryptography libraries implement the same algorithms (TLS, RSA, SHA1), and data manipulation libraries implement various formats (JSON, XML, YAML).
%Other domains, such as collection libraries, or logging, might be less formally specified. Yet, previous work shows that they can still be substituted \cite{slf4j,shacham2009chameleon}.

This natural diversity of library implementations for specific features has been harnessed in previous work.  Koopman and DeVale studied the diversity of POSIX implementations and use it for reliability \cite{koopman1999comparing}. Shacham and colleagues exploit the diversity of collection libraries \cite{shacham2009chameleon} to optimize  the applications depending on them. Sondhi~\cite{sondhi2021mining} reuses the  test cases of similar libraries. Our work is the first to propose to harness diverse libraries to generate variant applications.
%\david{Here I am expecting a "However, none of the ..."}
%There is no "however" as this section only states the existence of diversity in libraries available in software repository. Next section presents the risks induced by library usages.

\subsection{Library Monoculture}
\label{sec:monoculture}
%Make the argument that 1 out of 4 library in maven central actual depends on at least of the top 100 library, (and in average 2.5)

While there exist a diversity of library implementations, the actual usages are skewed on a small subset of them. Applications massively depend on a handful of packages, while most other packages are rarely depended upon~\cite{decan19}.  Library usages in a software repository follow a power law~\cite{louridas2008power}.
Zerouali and Mens~\cite{zerouali2017analyzing} found that $97\%$ of the projects using Maven on GitHub declared a dependency towards \texttt{junit}. Meanwhile, the second most popular test library, \texttt{TestNG} is only used by $11.9\%$ of the projects of their dataset (most of them also using \texttt{junit}). This unequal distribution of usages within a library reservoir is far from an exception. For example, \textit{\url{mvnrepository.com}} lists \np{11628} usages within Maven Central of \texttt{Apache HttpClient}, the most popular HTTP client library and only \np{6604} for the second most popular one \texttt{OkHttp}.
The large number of applications that rely on the same popular libraries is evidence that a new form of software monoculture~\cite{birman2009monoculture,goth2003addressing} emerges in software repositories. 

The concentration of usages magnifies the risks of software supply chain attacks that target the popular libraries shared across a large number of applications.
Meanwhile, there exist alternative implementations that provide the same features as these popular libraries. In this work, we propose to harness this diversity of implementations in order to mitigate the risk of  having one vulnerable library used by all applications.
Today, the discrepancy among alternative APIs and the entanglement between application and library code prevent a smooth, automatic replacement of one library by another. Some previous work lay the ground for this. In the Java ecosystem, Java Specification Requests~\cite{jsr} are formal documents to standardize the APIs of the different libraries of a specific domain. However, the general case is that the multiple libraries which implement similar features are not inter-changeable, as they have different APIs.
On the client side, there is a rich literature on API migration~\cite{alrubayeM019,balaban2005refactoring} that describes how to modify clients calls to APIs to adapt them to alternative libraries. However, none of these approaches allow a fully automated migration of existing clients from a library to an alternative. 

State-of-the-art techniques for API specification and migration do not support the automatic synthesis of diversified applications based on the alternative library implementations available in software repositories.In the next section, we discuss one key conceptual challenge that prevents this form of diversification in the supply chain.

\subsection{Client-library Entanglement}
\label{sec:entanglement}

%The following paragraph illustrates why, currently, different libraries offering similar features are not trivially replaceable, in general, for their clients. First, we describe how developers add dependencies in their project, and how it leads to the project being entangled with the dependency's API.

When the developers of an application identify the need for a specific feature,  they will usually find several libraries that provide the desired feature. 
For example, the \textit{reservoir} for HTTP client libraries in Maven Central contains libraries such as \texttt{Apache HttpClient}, \texttt{OkHttp}, or \texttt{Jetty client} \footnote{\url{https://mvnrepository.com/open-source/http-clients}}.
Yet, to use the desired feature, a developer needs to declare, in the configuration file of its package manager, a dependency towards one specific library. For example, a developer will not declare a dependency towards an abstract HTTP client, but towards the specific library \texttt{Apache HttpClient} version \texttt{4.5.13}. The package manager will then automatically fetch the library and bundle it with the rest of the application.
Once the chosen library is fetched, the desired feature can be used through direct calls to the library's API. For example, in the case of \texttt{Apache HttpClient}, it may be instantiating an object from the class \texttt{CloseableHttpClient} and calling its method \texttt{execute} to execute an HTTP query.

Application developers  might need an abstract feature regardless of what library implements it. Yet, to use this feature, they need to choose \textit{a specific library} and make their code \textit{entangled} with its API. Meanwhile, alternative libraries might propose very different APIs, with different abstractions. Methods might take a different number of parameters, or parameters of different types. This makes migrations from a library to an alternative potentially time costly \cite{Alrubaye18}.
Furthermore, migration requires good knowledge of the former library's API, the new library's API and the ability to modify the code of the client. 
This problem intensifies in the context of transitive dependencies, i.e., third-party libraries depended upon by a client's library. 
For example, \texttt{Apache HttpClient}, in its version \texttt{4.5.13}, depends on other libraries such as \texttt{commons-codec} and \texttt{commons-logging}. This means that a project that depends on \texttt{Apache HttpClient} also  depends on these two libraries. In this case, the developer of such a project typically does not have the possibility of modifying the source code of \texttt{Apache HttpClient}, hence performing a migration from, \texttt{commons-codec} to an alternative library is unpractical.

The entanglement between application code and APIs,  the diversity of APIs in a reservoir and the complexity of dependency trees are key challenges that we need to address in order to automate library substitution.

\section{Automatic diversification in the software supply chain}
\label{sec:concept}

In this work, we aim at letting developers benefit from the natural diversity of libraries in a given domain. Our goal is to build diverse application variants, with the same functionalities, but diverse supply chains. 
Given an application that depends on a specific library, we propose a systematic way to substitute this library, without modifying the source code of either the application nor the library. %In \autoref{sec:entanglement} we discuss why current practice of dependency reuse, prevents from benefiting from a diversity of suppliers. 
%\autoref{sec:adapters} introduces our proposition to address current limitations by decoupling a library implementation from its API. \autoref{sec:archi} describes an architecture for library substitution, based on a 3-tier delegation approach, and a careful curation of APIs to limit development effort. \autoref{sec:eval-goals} summarizes the assessment criteria of a Library Substitution Framework.

\subsection{Generating a Population of Variants}
\label{sec:multivariant}

A software supply chain attack that targets one single library has the potential to disrupt all the  instances of an application that depends on the library. 
In this work, we propose to mitigate this risk through supply chain diversification. We aim at synthesizing a population of application variants which instances are different enough, so they cannot be attacked in the same way. This is achieved because these instances do not depend on the same libraries implementations, hence are not sensitive to the same supply chain attacks.

For a given application that depends on a specific library, we propose to generate as many variants as there exist alternative libraries providing similar features. For example, from an application depending on \texttt{Apache HTTPClient}, we propose to generate a population of variants, one depending on \texttt{OkHTTP}, one depending on \texttt{Eclipse Jetty}, one depending on \texttt{Jode HTTP}, etc. This population would include one variant per alternative library that \texttt{Apache HTTPClient} can be substituted with, without breaking the functionalities of the original application.

To produce these variants, we propose a generic architecture for library substitution. A Library Substitution Framework is specialized for a reservoir, and supports the generation of variants, each based on an alternative library of the reservoir.
Overall, the goal of such a framework is to maximize the number of variants that can be created from a single client application, depending on a library of the reservoir.

However, substituting a library by an alternative raises significant challenges because of client-library entanglement and API discrepancies. In the following sections, we propose a general architecture that addresses these challenges.

%What to do with the population

\subsection{Library Adapters}
\label{sec:adapters}

\begin{figure}[ht]
    \centering
    \includegraphics[width=0.9\columnwidth]{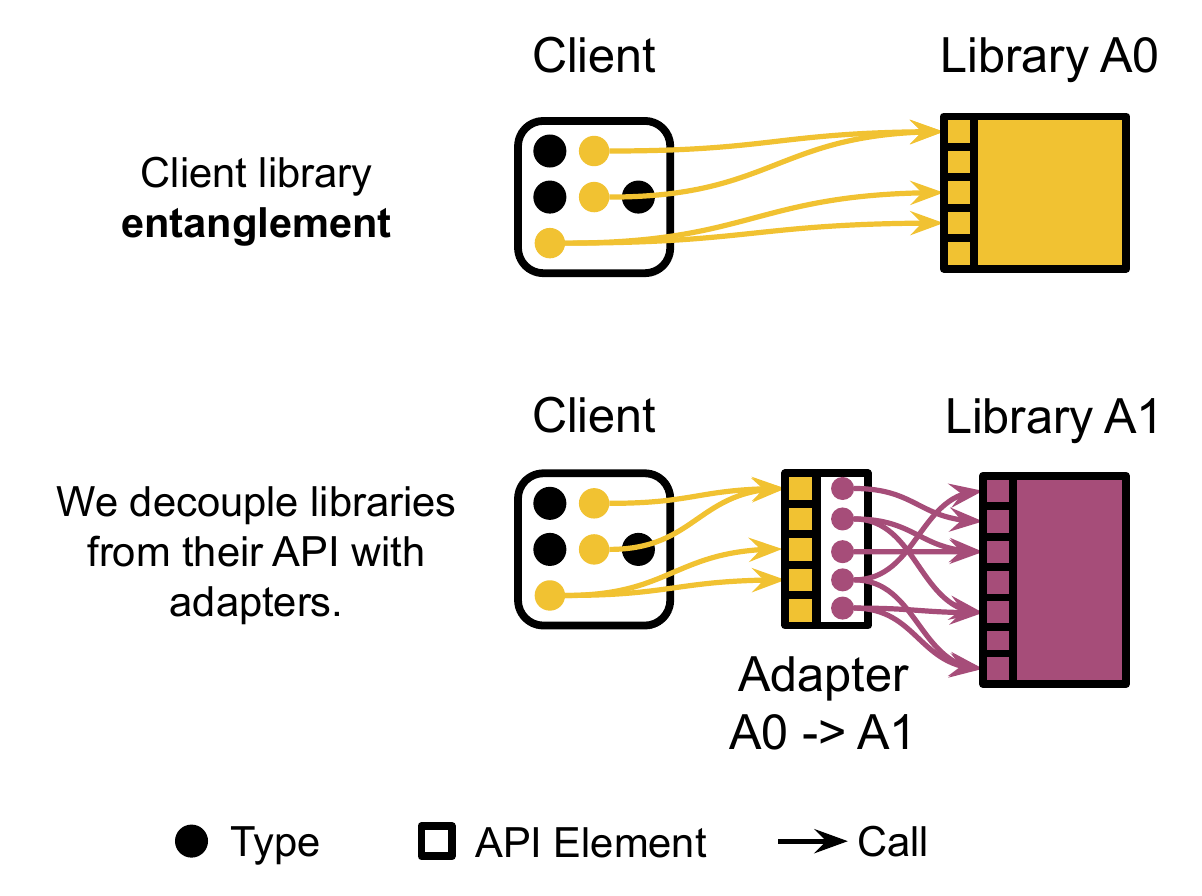}
    \caption{To support library substitution, we propose to define adapters between the API of a third-party library and other libraries from the same reservoir. Untangling 'behind' the API allows us to substitute similar libraries with no modification in the client application code.}
    \label{fig:adapter}
\end{figure}

%Library are here to stay.
%If we cannot avoid depending on external code, we can at least gives the opportunity
The key challenge for automatic library substitution lies in the tight entanglement between an application and the APIs of its third-party libraries (top part of \autoref{fig:adapter}). To support substituting a library by another in the same reservoir, we propose to introduce one level of indirection between an API and its  implementation. Keeping the original API allows us to support library substitution without modifying the code of the client application.
The level of indirection between the API and an implementation is a piece of software that binds the API  elements to elements of the implementation. We call this piece of software an \textit{adapter}. 
\autoref{fig:adapter} illustrates the concept. It represents a client that depends on a library A0. The client includes invocations to the API of A0. Hence,  replacing A0 by A1 is not possible, as A1 exposes a  different API. An \emph{adapter} from A0 to A1 exposes the same API as A0, and it implements wrappers for all API elements of A0 that adapt the calls towards A1.
For example, an adapter from \texttt{Apache HttpClient} to \texttt{OkHTTP} would provide the API of \texttt{Apache HttpClient} (with types such as \texttt{CloseableHttpClient}) but calls to this API would be translated to calls to the API of \texttt{OkHTTP}.

\subsection{Library Substitution Framework Architecture}
\label{sec:archi}

\begin{figure*}[ht]
    \centering
    \includegraphics[width=0.9\textwidth]{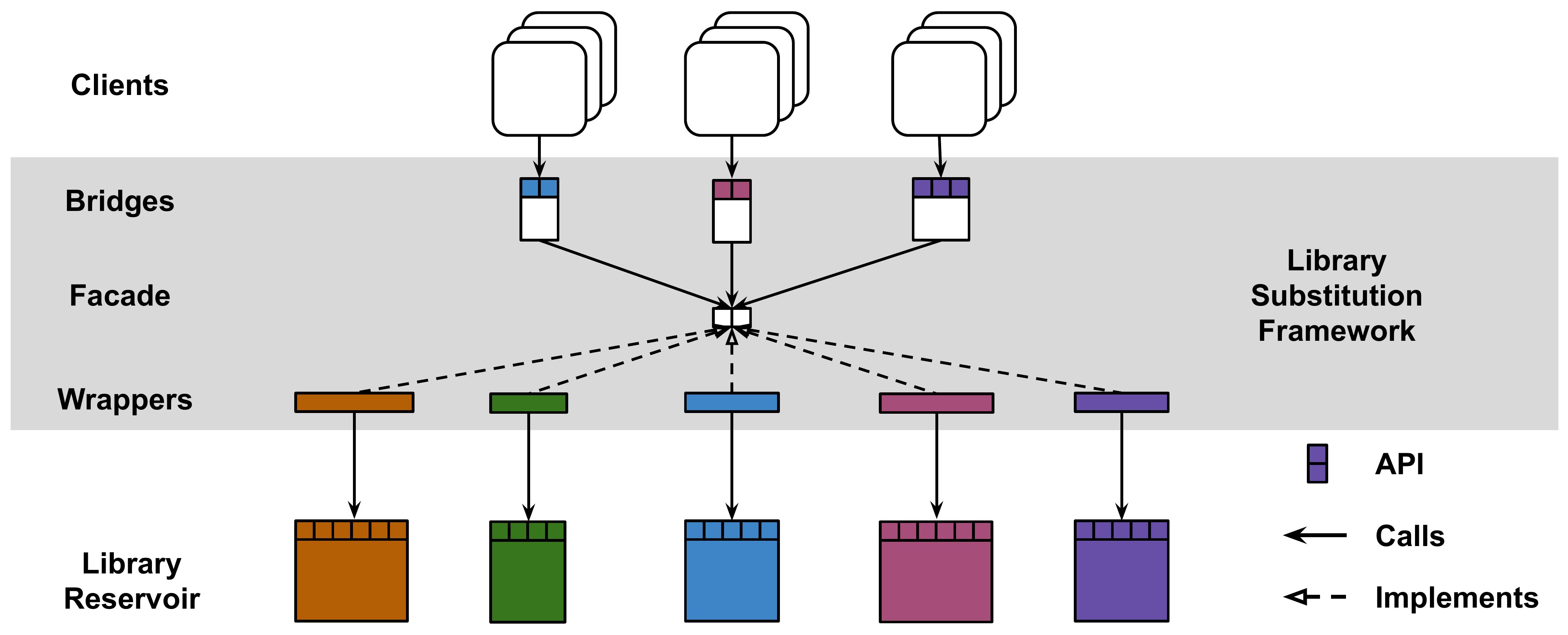}
    \caption{Generic architecture for a Library Substitution Framework that targets a set of applications and a library reservoir. Bridges abstract the API used by the clients from the concrete implementation; the facade defines a common API for all libraries; a wrapper maps the abstract API to the concrete functions of an existing library of the reservoir.}
    \label{fig:gen-wrapper-archi}
\end{figure*}

In this section, we refine the high-level concept of \emph{adapter} into a generic architecture to build a Library Substitution Framework. In particular, this architecture aims at limiting the development effort related to the high number of API elements to adapt. %Yet, this need must be put into balance with the need to still support library substitution for a majority of client applications depending on libraries of a given reservoir.
%The diversification of all clients of all the libraries in a reservoir presents some significant development effort, as it requires a great number of adapters. In this section, we discuss how to curate the APIs to be adapted, in order to bring the development effort down to a realistic amount.
A naive approach would implement an adapter for each pair of libraries in a given reservoir, with each adapter implementing the full API of the library being adapted. For a reservoir containing $n$ libraries, the number of API element implementations that the framework needs to provide is computed with:
%\begin{theorem}
%\label{eq:original}
%\textbf{Number of API element adaptations.}
\begin{equation}
    \sum_{i=0}^{n} \#Version_{API\_i} \times Size_{API\_i} \times (n - 1)
\end{equation}
%\end{theorem}
%\nicolas{Remove Equation, and cite as (1)}

Indeed, for each version ($\#Version_{API\_i}$) of each of the $n$ libraries in the reservoir,  every API element ($Size_{API\_i}$) needs to be re-written with each of the $n-1$  other libraries of the reservoir. The amount of code to be developed, tested, and maintained makes this naive approach impractical for large library reservoirs.

We leverage one key observation to limit the development effort for adapters: the distributions of usages among libraries in a reservoir\cite{harrandjson}, among versions\cite{soto2019emergence} and among the API members\cite{harrandCore}, are skewed towards a handful of popular elements. If the designers of a Library Substitution Framework accept to build a solution that is effective  for a majority of client applications in a given domain, instead of all of them, then, the complexity of the framework can be significantly reduced. We discuss each point in the following paragraphs.

\textit{Size of APIs:} The number of elements exposed by a library's API can be large. For example, among the $100$ most popular libraries of Maven Central, the median number of public types is $202$, the median number of methods $1,815$ and the median number of public fields $148$~\cite{harrandCore}. There might be some client applications that depend on these elements. Consequently, a Library Substitution Framework shall support all these API elements to provide diversity to all applications. However, the vast majority of clients rely on a small subset of the original APIs~\cite{harrandCore}. Consequently, framework designers can provide adapters that focus on the most popular subset of the original libraries' API.

\textit{Number of versions:} A library is released through multiple versions, and all versions are available on the online repositories. For example, in Maven Central, there are between $5$ and $200$ different versions for $50\%$ of the libraries \cite{soto2019emergence}. Each new version potentially changes the API of the library. To support all clients of all versions of a library, the adapter needs to support the union of the APIs of all versions. However, only a handful of versions are used by a majority of clients~\cite{soto2019emergence}. This implies that a Library Substitution Framework can focus on the most popular version of each library. Furthermore, the subset of popular API elements of a library does not necessarily change for every version~\cite{breakingbad}. Hence, the most popular version of an API may also be compatible with clients of other versions. 

\textit{Number of Alternative Libraries:} Developing adapters for each library in a reservoir can be challenging, in particular for a large reservoir. For example, \textit{\url{mvnrepository.com}} lists more than $80$ different HTTP clients libraries for the JVM, or more than $40$ different logging frameworks. Designers can focus on supporting a minority of popular libraries. Consequently, their Library Substitution Framework will not be usable for the few client applications that use the least popular libraries in the reservoir. 

In addition to the curation of the API to adapt, we propose a generic architecture for a Library Substitution Framework. In particular, this architecture addresses the problem of the number of possible substitutions.
The number of alternative libraries also increases quadratically the number of combinations of substitution possible.
To limit this quadratic explosion, we rely on the inversion of control pattern~\cite{inversion}. We implement adapters between the APIs used by the client applications and a common, abstract API that captures the core set of features shared by all libraries in the reservoir. We then implement adapters between this common API towards each library of the reservoir. Hence, the actual library can be changed seamlessly.
\autoref{fig:gen-wrapper-archi} describes the proposed three layers architecture.

\textbf{Bridge:} A bridge aims at abstracting the API of a library used by a client, without modifying the code of the client. To do so, a bridge presents to its clients the same API (or a subset) as the library it substitutes. The actual implementation of the bridge actually transfers method calls to objects implementing common interfaces contained in the facade. A bridge ensures the syntactic correctness of the substitution. To summarize, a bridge is an adapter from the API of the library it substitutes towards the API described in the facade.

\textbf{Facade:} A facade specifies a set of abstract operations that capture the core set of features that are provided by all libraries in the reservoir. It defines a common API that bridges use. It also defines a Factory that instantiates a concrete library that will provide these operations, providing the inversion of control mechanism~\cite{inversion}. 

\textbf{Wrapper:} A wrapper is an adapter from the API described in the facade towards an existing alternative library. It provides the concrete behavior for the operations specified in the facade. To be part of this architecture for substitution, the library must be extended with adapted classes that relate the operations of the facade with the concrete classes of the library. The more libraries naturally available in a reservoir, the more behavior diversity we can introduce for the client applications.

With this architecture, a client application can benefit from the alternative libraries by replacing, in the configuration file of its package manager, the original dependency by its corresponding bridge and the desired wrapper. The pair Bridge / Wrapper constitutes the adapter from the bridged library to the wrapped one.

The architecture of \autoref{fig:gen-wrapper-archi} combined with the API curating, greatly reduces, the number of API elements to be adapted, which can be revised from (1) to:

\begin{equation}
    \sum_{i=0}^{m} Size_{Most\_Used\_Elements(API\_i)} + \sum_{i=0}^{n} Size_{Facade}
\end{equation}
    with $m \ll n$
%This number goes from  \autoref{eq:original} to \autoref{eq:revised}.

First, only the most popular version of an API is selected, second, only the most popular elements of the  $m$ most popular APIs are adapted, and they are adapted only once thanks to the facade. Then, each API element of the facade is implemented once per library ($n$) in the reservoir. It is important to note that this approach scales linearly with the number of libraries $n$ in the reservoir, rather than quadratically. Furthermore, this architecture makes the addition of new libraries to the reservoir simpler to handle. Indeed, both the development of a new bridge and a new wrapper only require knowledge of the API of the new library and the facade.
%\begin{theorem}
%\label{eq:revised}
%\textbf{Revised number of API element adaptations.} 
%\end{theorem}

%\nicolas{Now Adapter is made out of Bridge + wrapper. PS remove sentence saying: "others call it a wrapper".}

%\todo{Add a paragraph to introduce the transition, ask questions}

\subsection{Assessment Criteria for Library Substitution Frameworks}
\label{sec:eval-goals}
%\nicolas{todo}
%To summarize, our propose architecture for Library Substitution Framework must answer two sets of  goals in tension:
%\begin{enumerate}
%    \item \textbf{Goal of diversity:} The architecture must provide a maximum of variant per client. Variants should exhibit behaviors as diverse as possible, yet equivalent.
%item \textbf{Goal of efficiency:} The architecture must minimize development cost. Yet, the architecture must maximize the number of clients for which a library can be substituted by an alternative.
%\end{enumerate}
To evaluate the relevance of the concept of Library Substitution Framework proposed in this work, as well as the architecture described in this section, we formulate 4 criteria, associated to 4 metrics

\textbf{Minimal number of API elements to adapt:} An important aspect of the architecture we propose in this work, lies in curating the APIs to adapt. The metric to observe, regarding that aspect, is the number of API elements that actually need to be adapted.

\textbf{Preservation of the behavior of the bridged libraries:} Given a bridged library, the bridge should behave as the original library, with all implementations in a reservoir. In order to assess the behavior preservation, we rely on the existing test suite of the library. We measure this with respect to the number of tests of a library that pass on its corresponding bridge coupled with each wrapper in the reservoir.

\textbf{Maximize the share of clients supported by the bridge APIs:} A bridged library does not adapt all API elements of the original library. Still, a Library Substitution Framework aims at supporting the diversification of a maximum number of applications. We measure the effectiveness of this trade-off with the share of clients that compiles when a library is substituted by its corresponding bridge.

\textbf{Maximize the number of behavior preserving variants per client:} The main goal of a Library Substitution Framework is to diversify the supply chain of the applications that depend on library of a reservoir. Meanwhile, diversification must preserve the original behavior of the application. Here, we count the number of application variants  that  pass the original test suite, after substituting the original dependency by one of the implementations in the reservoir.

\section{Experimental Protocol}
\label{sec:methodology}
%\nicolas{To do update}
%\benoit{Replaced 'swap' by 'substitute'; this needs to be checked globally}

%\benoit{check tenses, use only present}

%In this section, we describe our methodology to investigate \argo's impact on the clients of the four \json libraries for which we built bridges. \autoref{sec:rqs} presents our research questions, \autoref{sec:meth-dyn} describes the \placebo \json library, a library used to assess whether a test suite covers the usage of a \json library. In 
%\autoref{sec:methodo-rq1} we introduce our protocol for RQ1. In \autoref{sec:clients} we present how we create a dataset of Java projects depending on \json libraries, and in \autoref{sec:methodo-rq234} we detail the protocols to answer our remaining research questions.

The architecture, presented in the previous section, is generic and aims at being instantiated for specific library reservoirs.
To evaluate the feasibility, and the effectiveness of Library Substitution Framework, we implement a case study for the reservoir of Java \json libraries. 

JavaScript Object Notation, or \json, is a ubiquitous file and data exchange format. Since its creation in 1999, it has gained considerable popularity for a wide spectrum of activities such as web APIs \cite{tan2016service}, scientific computing \cite{millman2014developing}, data management \cite{LiuHMCLSSSAA20}, or gaming \cite{tost2021}. It is also used as a serialization format for Java objects.
%Serialization vulnerable
Serialization is known to be vulnerability prone~\cite{Peles15,Viega0000,Fingann20}, and JSON libraries are no exception~\cite{friday13json}. For example, \texttt{jackson-databind}, the most popular Java \json library is referenced in at least $68$ CVE~\cite{jacksonCVE}.
This makes JSON libraries good targets for supply chain attacks on third-party packages. Therefore, the reservoir of  JSON libraries \cite{harrandjson} is a relevant target to experiment the concept of Library Substitution Framework.

%We then split our research according to two axes. First, we evaluate the feasibility of a such a framework for \json libraries that preserves the common features of this reservoir. We implement a concrete framework named \argo.\footnote{Argo is Jason's mythical ship, and like Theseus' ship, replacing part of it raises the question of the ship remaining the same ship or not.} From \argo's development, we assess both the development effort, and the ability of the framework to provide library substitution that are behaviorally equivalent modulo test.
%Second, we investigate the impact of the framework on real open-source client applications, depending on \json libraries mined on GitHub. From this investigation, we assess how successful at generating a maximum of behaviorally equivalent variants for a maximum of clients.

\subsection{Research Questions}
\label{sec:rqs}

\newcommand{\RQone}{\textbf{RQ1. Can the architecture we propose for a Library Substitution Framework be implemented from a concrete reservoir of libraries?}}

\newcommand{\RQtwo}{\textbf{RQ2. Can the test suite of libraries, be curated and reused to test a Library Substitution Framework?}}
%\newcommand{\RQtwo}{\textbf{RQ2. Can the preservation of features of the libraries be tested by reusing their existing test suites?}}
% How diverse are the \json libary \todo{implementations} in the JVM ecosystem

\newcommand{\RQthree}{\textbf{RQ3. What is the share of clients for which the framework preserves the static semantics of the original dependency?}}
%RQ2. What is the share of clients that compile correctly with the bridge corresponding to their original  \json dependency?

\newcommand{\RQfour}{\textbf{RQ4. How many variants can we produce for each client, while preserving the original behavior?}}

\newcommand{\RQfive}{\textbf{RQ5. How can clients increase the diversity of their providers?}}

%\newcommand{\RQfive}{\textbf{RQ5. To what extent does the diversification of \json provider also diversify the observable behavior of \json clients?}}

%\benoit{the wrappers and implementations have a strong impact on behavior, why does the RQ focus on bridges?}
%\nicolas{Add two sentences to present first rq group}

In \autoref{sec:concept}, we introduced a general architecture for a Library Substitution Framework that supports the automatic build of variant application with diverse suppliers. We validate our approach for \json through 5 RQs: 2 that focus on validating the architecture and 3 that validate the support to build variants.

The first group of research question assesses the feasibility of implementing a Library Substitution Framework as described in \autoref{sec:archi}, and testing its features. To answer these questions, we focus on the \emph{libraries} of the reservoir.

\textbf{\RQone}

In this first research question, we implement \argo, a Library Substitution Framework for \json libraries. We follow the process of curating APIs and the general architecture described in \autoref{sec:archi}. We describe the design decisions such as the choices of bridged libraries, the choice of features to include in the facade, as well as the implementation of wrappers. We then discuss the impact of these choices on the number of API elements adapted in the framework.

\textbf{\RQtwo}
%\nicolas{Litteraly say that we reuse test of existing library to test the framework/bridges:
%Can we reuse test suite? Emphasize that we validate the test selection process...}

%To be useful, libraries of the reservoir targeted by a Library Substitution Framework, need to exhibit different behaviors. Yet, to be substitutable, these libraries must offer similar features.
%The goal of a Library Substitution Framework is to let client applications replace a library by an alternative offering an equivalent yet diverse behavior. 
%In this research question, we evaluate to what extent our concrete framework, \argo, does indeed offer an equivalent behavior compared to those of the original libraries. To do so, we rely on the existing test suite of the $4$ bridged libraries, and run a subset of their tests on the $20$ libraries of the reservoir through \argo's bridges and wrappers.
In this research question, we investigate to what extent we can reuse the test suites of the bridged libraries to validate a Library Substitution Framework.
%We focus on our concrete framework, \argo. 
We carefully curate the test suites of the $4$ libraries bridged in \argo. We then run these test suites on each of the $80$ combinations of bridges and wrappers available in the framework.
This reuse of tests across libraries of the same domain is in part similar to the work of Sondhi and colleagues
\cite{sondhi2021mining}.

\vspace{1em}
%\nicolas{Add two sentences to present second RQ group}
In research questions 3, 4 and 5, we assess the impact of a Library Substitution Framework on \emph{clients} of libraries from the targeted reservoir. These questions focus on the impact of \argo on the clients of \json libraries.

\textbf{\RQthree} In this research question, we assess which clients compile correctly when replacing the original library by the corresponding bridge. This means that all API members used by a client do have a substitution provided by \argo.  We also investigate cases where this condition is not met. This allows us to study the consequences of the trade\-off, described in \autoref{sec:archi}, between the size of the APIs bridged and the number of supported clients.

\textbf{\RQfour} This research question is at the core of our work. Here, we measure  the number of \json libraries we can automatically substitute while preserving the original behavior of the clients, as specified by their test suites. The larger number of libraries we can substitute, the more diversity of providers  \argo can support in the supply chain of the clients.

\textbf{\RQfive}
When the test suite of a client fails, the substitution of the library it uses, it gives us indications on what makes clients tied to a specific \json library. By analyzing these test failures, we draw general principles that make clients less coupled with a library and more prone to diversification.

\subsection{Datasets}

In this section we present the two datasets we build to study our two groups of research questions. First, we collect a reservoir of Java \json library to build a concrete Library Substitution Framework, described in \autoref{sec:json-family}. Second, we gather a set of clients of the most popular libraries in this reservoir, described in \autoref{sec:clients}, to study the impact of our framework.

\subsubsection{JSON Library Reservoir}
\label{sec:json-family}
%\nicolas{Move?}

%In this section, we describe our implementation of the architecture proposed in \autoref{sec:concept} for the case of Java \json libraries, which we call \argo \footnote{Argo is Jason's mythical ship, and like Theseus' ship, replacing part of it raises the question of whether the ship remains the same ship or not.}.

%\input{tables/libraries_new}
\begin{table}[ht]
\centering
\begin{tabular}{@{}lrrrrr@{}}
  \toprule
\textsc{Library} & \textsc{\# Commits} & \textsc{\# Stars} & \textsc{Version} & \textsc{Bridged}\\ 
  \midrule
  gson & 1485 & 18.8k & 2.8.5 & \checkmark \\ 
  jackson & 7382 & 2.7k & 2.12.0-rc2 & \checkmark \\ 
  json & 841 & 3.7k & 20201115 & \checkmark \\ 
  json-simple & - & 594 & 1.1.1 & \checkmark \\ 
  \midrule
  cookjson & 116 & 3 & 1.0.2 &  \\ 
  corn & - & - & 1.0.8 &  \\ 
  fastjson & 3793 & 1.4k & 1.2.75 &  \\ 
  flexjson & - & - & 3.3 &  \\ 
  genson & 395 & 193 & 1.6 &  \\ 
  jjson & 216 & 12 & 0.1.7 &  \\ 
  johnzon & 780 & - & 1.1.8 &  \\ 
  json-argo & - & - & 5.13 &  \\ 
  json-io & 1040 & 268 & 4.12.0 &  \\ 
  json-lib & - & - & 3.0.1 &  \\ 
  json-util & 464 & 48 & 1.10.4-java7 &  \\ 
  jsonij & 348 & - & 0.3.1 &  \\ 
  jsonp & 530 & 75 & 2.0.0 &  \\ 
  mjson & 79 & 67 & 1.4.0 &  \\ 
  progbase & - & - & 0.4.0 &  \\ 
  sojo & - & - & 1.0.13 &  \\ 
   \bottomrule 
\end{tabular}
    \vspace{0.5em}
    \caption{Description of the \json libraries studied by \textit{Harrand et al.}~\cite{harrandjson}}
    \label{tab:libs-api}
\end{table}

Our reservoir of Java \json libraries includes $20$ libraries available on Maven Central, all providing (i) \json parsing features, and (ii) serialization to \json text. These libraries have been designed by different people with different motivations.
We have previously analyzed the behavior of this reservoir in depth~\cite{harrandjson}. In our previous work, we make three points that are of interest for this work: (i) these libraries make a large variety of design decisions to represent \json data; 
(ii) these libraries exhibit a large diversity of behaviors regarding what \json inputs they consider as well-formed or ill-formed;
(iii) there is more diversity of behaviors observed on \json inputs that do not conform to the standard grammar~\cite{rfc8259}.

\autoref{tab:libs-api} describes our  reservoir of \json libraries. 
%Column \textsc{\#Version} gives the number of versions on Maven Central, for each library, as listed by \textit{\url{mvnrepository.com}}. Note that this number is a low approximation, as some libraries changed their names in the past. 
Column \textsc{\#Commits} indicates the number of commits in the source repository of the library, when available.
Column \textsc{\#Stars} gives the number of stars each library has, when hosted on GitHub.
As described in \autoref{sec:archi}, supporting all versions of all libraries, is a large effort. Hence, we focus on the most popular version of each library.
Column \textsc{Version} shows the version selected to build \argo. 
%Column \textsc{API Size} indicates the number of public elements (Constructors, fields, and methods) exposed by the API of the chosen version of each library. These elements can be called by clients of these libraries.
Finally, we select the API of $4$ of the most popular libraries, \jackson, \gson, \orgjson, \jsonsimple for which we develop bridges. We also collect their respective test suites to assess the preservation of their features by \argo.

%\textbf{\placebo \json library: } Both group of research questions in these papers involve reusing test suites to evaluate if the behavior of our framework is satisfactory. Meanwhile, a major threat to validity when assessing the impact of a library substitution with test suites, lies in the difficulty of knowing whether the library is covered by the tests. First, previous work~\cite{sotobloat,harrandCore} demonstrate that roughly $40\%$ of libraries declared by maven projects are not used at all. This phenomenon is known as bloat. Secondly, even when a library is used by a software project, the test suite of the project might not cover the portion of software that actually calls the library's API.
\textbf{\placebo \json library: } We want to verify that, when we substitute a library by another, the new library, provided by the framework, is actually covered by the test suite. To do so, we add another wrapper to the $20$ existing libraries.
\placebo is a special wrapper. It contains classes that implement the complete interfaces of the facade, and all of its methods directly throw an exception when invoked. Hence, when a \json library is substituted with \placebo, two things can happen: (i) the execution of the test suite triggers an exception from \placebo and fails; (ii) or the test suite passes without triggering any exception. If no exception is triggered, we conclude that the test suite does not cover the usage of the \json library. Hence, the test suite cannot be used to assess if the library substitution was successful.
If an exception is triggered, the \json library usage is covered by the test suite.

%Columns \textsc{\#API\_Types} and \textsc{\#API\_Elements} indicates respectively the number of public types and public elements (Constructors, fields, and methods) exposed by the library API. These elements can be called by clients of these libraries.
%Column \textsc{\#API\_Bridge} shows the number of API elements adapted in the bridges. These elements' code is rewritten using calls to elements of \argo's facade.
%These numbers are provided for the subset of the most popular libraries for which  we have chosen to implement a bridge.
%Column \textsc{\#API\_Wrapper} indicates the number of API elements of the original libraries used in \argo's wrappers.
%In the following section, we discuss these numbers with more details.

\subsubsection{Client of JSON Libraries}
\label{sec:clients}

In this section, we describe the methodology that we follow to construct the dataset of open-source Maven Java projects that use JSON libraries.
We choose open-source projects in order to have a large diversity of library usages, and to study the impact of library substitution.
%, and shows the ability of Argo to introduce diversity in JSON library usage. 
We want open-source projects that have a reproducible build, use a \json library, and have reproducible tests that do cover the \json library in question.

The construction of this benchmark is composed of $4$ main steps. 
(i) First, we identify the \np{\nbTotalProject} Java projects on GitHub that have at least five stars. We use the stars as an indicator of interest. 
(ii) Second, we select \ChartSmall[q]{54703}{\nbTotalProject} Maven projects. 
We choose Maven projects to be able to build and execute the tests of the projects automatically.
%We identify them by listing all the files for each project and consider the projects that have at least a Maven build configuration file, (i.e., \texttt{pom.xml}).
(iii) Third, we parse each Maven build configuration file to identify the projects that use any version of a JSON library.
During this step, we identify \ChartSmall[q]{12012}{54703} projects.
(iv) As the fourth step, we execute three times the test suite of all the projects, as a sanity check to filter out projects that we cannot build and the projects with flaky tests.
We keep the projects that have at least one test and have all the tests passing: \ChartSmall[q]{1927}{12012} projects passed this verification.

This gives us \np{1927} client projects that declare a dependency towards a \json library and that can build.
We run an additional check to verify that the test suite of the client covers the usage of the \json library. Indeed, a test suite that does not cover the \json library, would pass regardless of whether our substitution is satisfactory or not. We substitute the \json library of these clients by the corresponding bridge and the \placebo implementation. If the test suite of the client passes, it means that the client does not actually need \gson to run. In that case, we remove the client from the dataset.

Our final dataset consists  of $368$ client projects distributed as shown in \autoref{tab:cli}. Column \textsc{\#Client} indicates the number of clients for each library, \textsc{\#Test} shows the combined number of tests provided by their test suite, and \textsc{\#JSON Test} gives the number of these tests that  cover the usage of a \json library. This number is obtained by checking the number of test failures with the \placebo wrapper.

\begin{table}[ht]
\centering
\begin{tabular}{@{}lrrr@{}}
  \hline
 \textsc{Library} & \textsc{\#Client} & \textsc{\#Test} & \textsc{\#JSON Test} \\ 
  \hline
 jackson-databind & 128 & 636,664 & 127,054\\
 gson & 122 & 201,576 & 6,891\\ 
 json &  84 & 84,030 & 10,835\\ 
 json-simple &  34 & 37,804 & 1,065\\ 
   \hline
 Total &  368 & 960,074 & 145,845\\ 
   \hline
\end{tabular}
\vspace{0.5em}
\caption{\json library clients studied}
\label{tab:cli}
\end{table}

\subsection{Experimental Protocol}
\subsubsection{Library Substitution Framework design and test (RQs 1 \& 2)}
\label{sec:methodo-rq1}

\begin{figure}[ht]
    \centering
    % https://drive.google.com/file/d/1C-pRmI-ADpRRsoFaKiARfTwL0Yi6Ye2S/view?usp=sharing
    %\includegraphics[width=0.5\textwidth]{figures/Library_experiments.pdf}
    \includegraphics[width=\columnwidth]{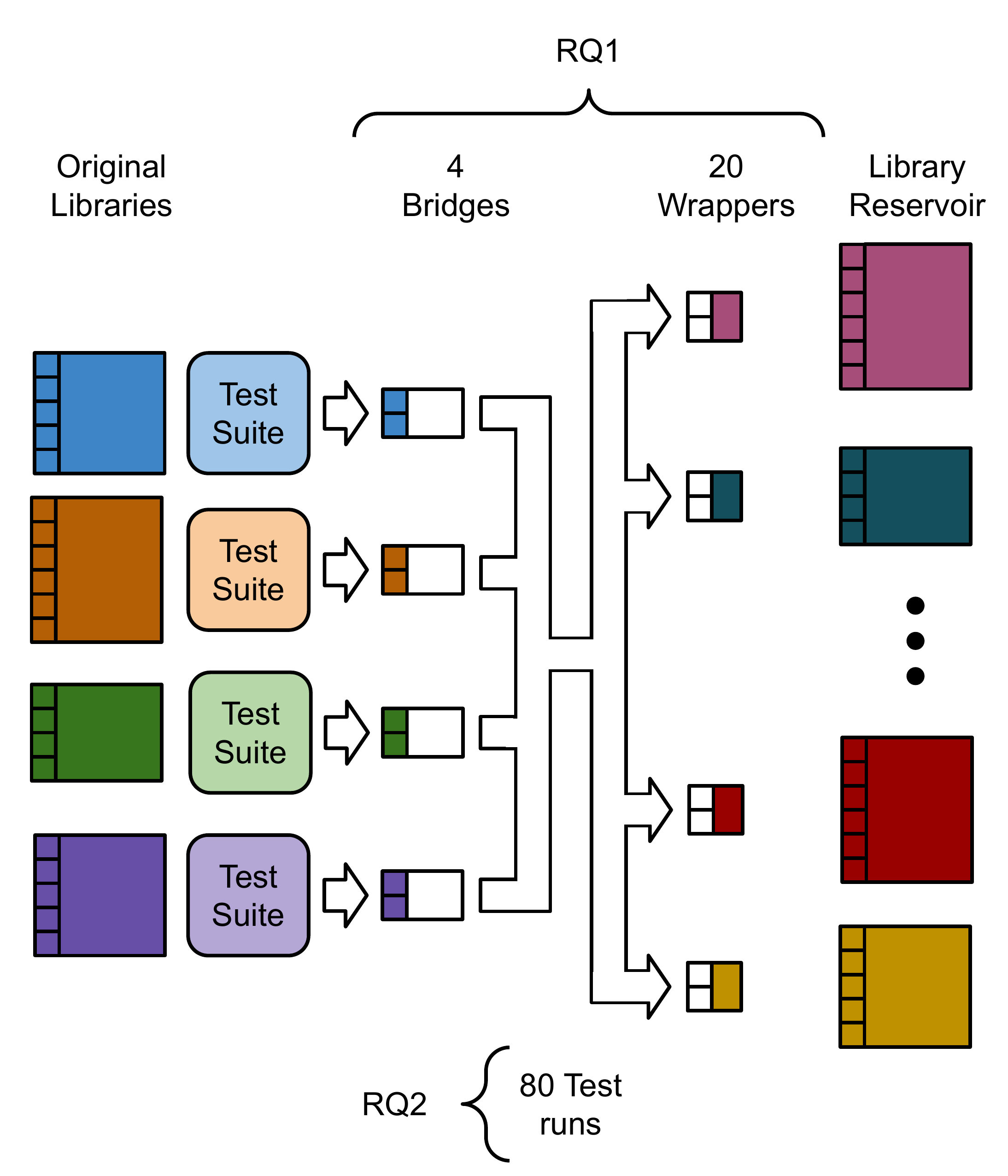}
    \caption{Design (RQ1) and test (RQ2) of \argo, our concrete Library Substitution Framework for the \json library reservoir. In RQ1 we design and implement the facade,  $4$ bridges, and $20$ wrappers. In RQ2, we curate and reuse the test suites of the $4$ bridged libraries, to run them on all combinations of bridges and wrappers.}
    \label{fig:lib-exp}
\end{figure}

The goal of our two first research questions is to assess the feasibility of implementing and testing the architecture described in \autoref{sec:archi}. \autoref{fig:lib-exp} gives an overview of our process to answer these questions.%We evaluate if a Framework can be 

\textbf{RQ1 protocol:} To answer the first question, we implement \argo, a Library Substitution Framework for the \json library reservoir described in \autoref{sec:json-family}. In order to bring down the development effort, we follow the process of API curation described in  \autoref{sec:archi}. We focus on the most used version of each library of the reservoir. We develop a bridge for $4$ of the most popular libraries: \jackson, \gson, \orgjson and \jsonsimple. In these bridges, we only adapt the most used API elements that correspond to the features shared by all libraries of the reservoir. We assess the popularity of API elements by static analysis of the clients' dataset described in \autoref{sec:clients}. To perform this static analysis, we reuse the tool develop for a previous study~\cite{harrandCore}.

We define a common API that describes the common features, which we package as the facade of our framework. These features are \json text parsing, \json type representation as Java Object, and data serialization as \json text. Finally, we write a wrapper for each of the $20$ libraries of the reservoir. We then evaluate this API curating process by discussing the number of API elements adapted compared to the total number of API elements in the reservoir.

\textbf{RQ2 protocol:} Once \argo is implemented, we run the test suite of the \nbsrcjson bridged libraries against their corresponding bridge in order to assess what functionalities of the \nbtargetjson libraries are preserved. %For example, we run the original test suite of \gson on the bridge \texttt{gson-over-argo}, which adapts \gson's API to \argo's facade. 
This assessment is based on a two-step protocol: (1) First, we curate the original test suite to select a subset of the original test suite that is relevant to test \argo; (2) we run this curated test suite on the corresponding bridge and each of the \nbtargetjson wrappers supported by \argo.

\begin{figure}[ht]
    \centering
    \includegraphics[width=\columnwidth]{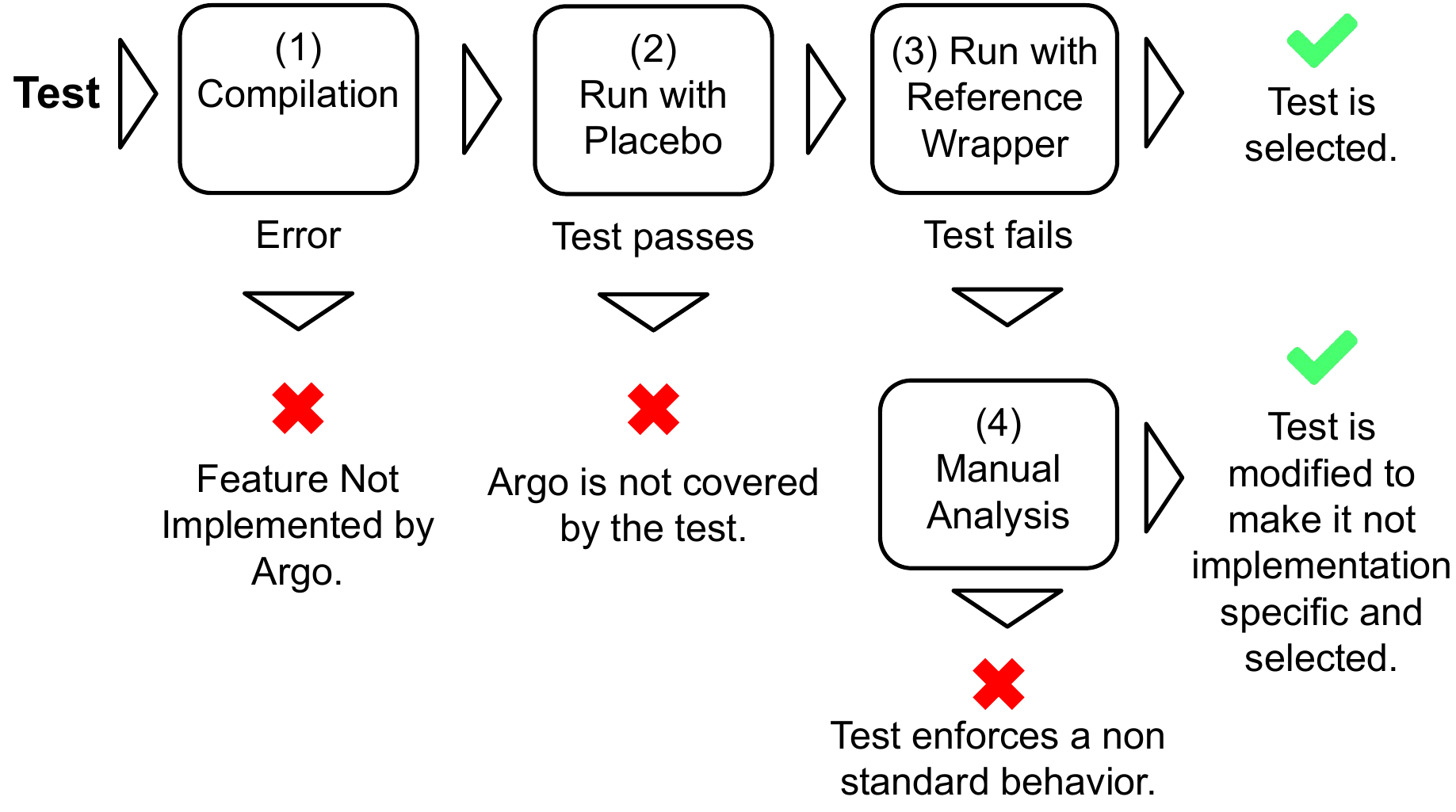}
    \caption{Protocol to select tests relevant to Library Substitution Framework validation from the test suite of the bridged libraries. Tests need to cover adapted API elements, while not enforcing a non-standard behavior.}
    \label{fig:test-selection}
\end{figure}
%\nicolas{may be remove fig3}
%\nicolas{may be add back a figure that shows that we are reusing test suite on bridges}

%The first step of the protocol consists in curating the original test suite. %This consists in selecting  the tests that cover features that are implemented in \argo. 
First, we follow the test selection process described in \autoref{fig:test-selection}.
Given one  original \json library, we extract its test suite and add a dependency to the corresponding bridge and implementation. For example, we take the test suite of \orgjson , and add a dependency towards \texttt{json-over-argo} and  \texttt{argo-json}. Hence, when a test case refers to a class from the original library, the corresponding class from the bridge is loaded instead. (1) First, we compile the test suite. Some test cases do not compile because the bridge does not implement the feature targeted by the test. We remove those tests. Then, we run the test suite on the bridge with the \placebo wrapper (2). We comment-out tests that pass despite the wrapper being empty and label them as \textsc{PLACEBO}. 
At the end of this test suite execution, we obtain a subset of the original test suite, composed exclusively of the test cases and assertions that assess the behavior of the bridge.

We execute the test cases that are not labelled \textsc{PLACEBO}, with the bridge and the wrapper that corresponds to the original library (3). We select all the tests that pass on this implementation.  If some test cases fail, we manually analyze them to determine whether to modify them or to remove them (4). We remove tests that specify a behavior stricter than what is described in RFC 8259~\cite{rfc8259}. 
The tests we modify are adapted in two ways. One is to relax assertions that assess strict equality of \json strings to make them tolerant to equivalent \json strings: make \json text equality check tolerant to white spaces or \json object tolerant to key reordering. 
The second is the relaxation of the number comparisons to make them tolerant to different types (removing unchecked casts, making floating-point comparison modulo epsilon, etc.).
As a sanity check, we re-run the whole selected test suite to make sure that every test fails on \placebo and succeeds on the reference wrapper.

The second main step of the protocol for RQ2 consists in running the selected test cases with every other library, to determine how much of the original behavior is preserved by the variants present in \argo. %For example, this means running the selected test from \gson's test suite on \gson's bridge with each of the $20$ wrappers available in \argo. Hence, we can observe which of those test passes or fails on each of the $20$ libraries. 
We then analyze the reasons for eventual failures.

\subsubsection{Client Experiments (RQs 3, 4 \& 5)}
\label{sec:methodo-rq234}

%Our second research axis regards the impact of a Library Substitution Framework on clients of the libraries of the reservoir targeted by the framework. We assess the consequences of our architectural choices on the share of client for which library can be substituted. We also evaluate the amount of variants that can be produced for each client. %Finally we investigate test failures to draw principles

\begin{figure}[ht]
    \centering
    %\includegraphics[width=\columnwidth]{figures/Test_Selection.pdf}
    % https://drive.google.com/file/d/1C-pRmI-ADpRRsoFaKiARfTwL0Yi6Ye2S/view?usp=sharing
    %\includegraphics[width=0.5\textwidth]{figures/Client_experiments_2.pdf}
    \includegraphics[width=\columnwidth]{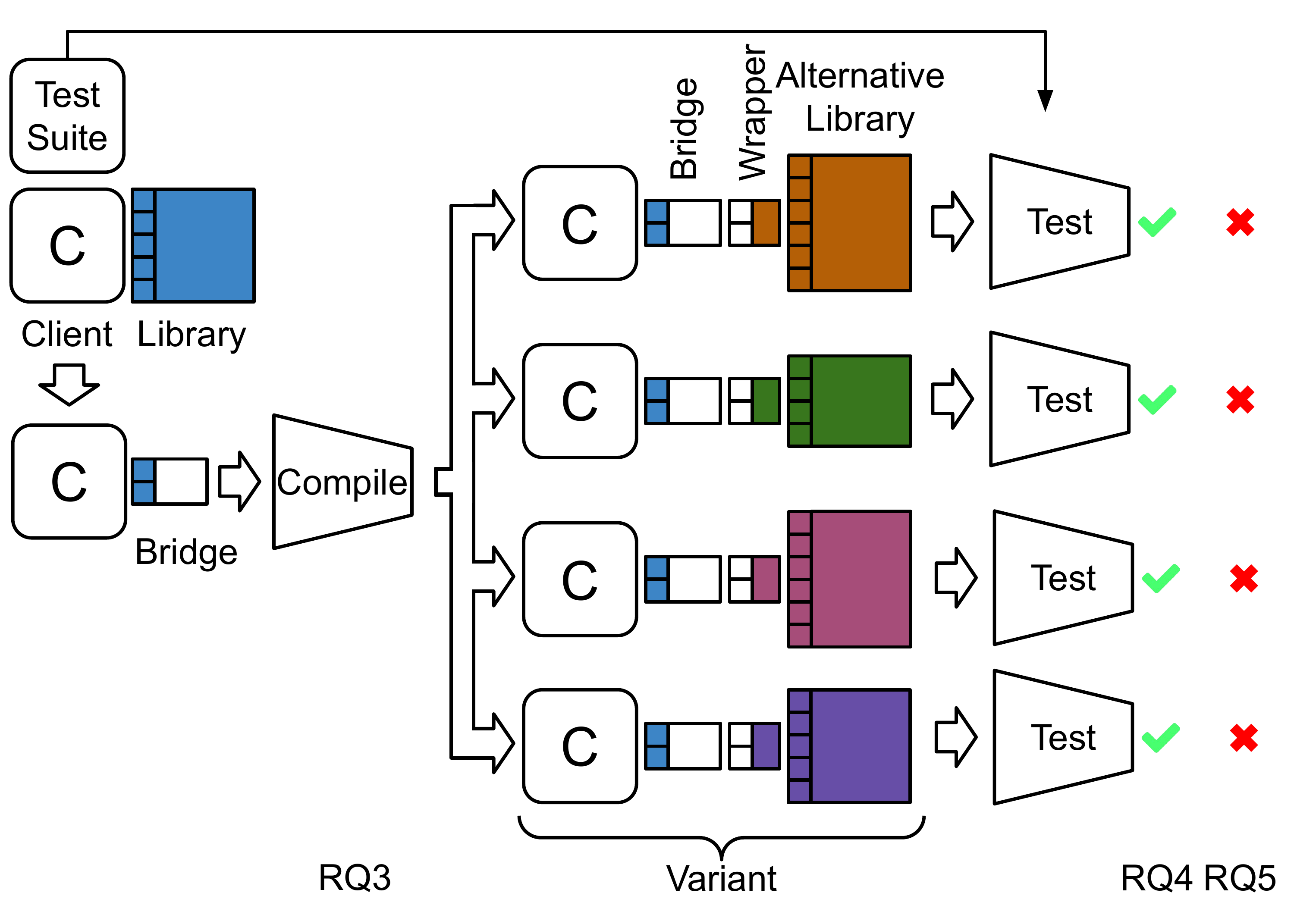}
    \caption{The flow of experiments on \json libraries' clients.\\
    For each client, we substitute the original library by its corresponding bridge. We, then, compile the client to detect potential missing API elements from the bridge (RQ3).
    If compilation succeeds, we generate 20 variants per client, one per library in the reservoir. We run the clients' original test suites on each variant. We then analyze test successes (RQ4) and failures (RQ5).}
    \label{fig:client-exp}
\end{figure}

Our second set of RQs assesses whether we can generate application variants through library substitution. To answer them, we substitute the \json library, used by the clients of the dataset described \autoref{sec:clients}, by alternatives provided by \argo. \autoref{fig:client-exp} describes the flow of our experiment. For each client, we substitute the \json library and replace it by the corresponding bridge.
Since all of our clients are Maven project, to perform substitutions, we edit the \textit{pom.xml} files of clients.
This allows us to alter the classpath of the client both during compilation and test execution.

%\nicolas{To fix: We attempt to compile this modified client, to detect static errors. For each client, we then add to their ach of the $20$ wrapper developed in \argo. This yields us $20$ potential variants, on which we run the original test suite.}
%RQ3 aims at measuring the proportion of clients that compile successfully when using \argo. To assess this success rate,
\textbf{RQ3 protocol:} We analyze the result of compiling each client, where the original \json library has been substituted by the corresponding bridge. 
Our dataset only contains client projects for which we can reproduce the build. Consequently,
a build failure with the bridge indicates that there are static differences between the part of the API of the original library used by the client and its \argo counterpart.
By contrast, when no compilation error occurs, it means that all API members statically called by the clients are exposed by the bridge. 

%\benoit{split into protocol for RQ3 and for RQ4; elaborate on what aspects you analyze in each RQ}
\textbf{RQ4 protocol:} In RQ4, we measure the number of behaviorally equivalent variants we can generate through library substitution.
We focus on clients for which the substitution of the library, by its corresponding bridge, leads to successful compilation. For each client, we generate $20$ variants, one per library in the reservoir described in \autoref{sec:json-family}. A variant is composed of the original client, a bridge, a wrapper and one alternative library.
We run the test suite of these clients on each of the $20$ variants. 
We assess if variants are behaviorally equivalent, by checking whether the client's test suite passes, without test failures, on them. 
Our dataset only contains clients for which the test suite covers the usage of the original \json library, as a substitution with the \placebo wrapper makes their test suite fail. Hence, we know that the substituted library is covered by at least one test in the client's test suite.
We then count for each client the number of behaviorally equivalent variants. 
This number of variants is the number of alternatives that our Library Substitution Framework provides for a given client.

%We focus on whether the original test suite of our clients still passes on each of the $20$ variants generated with \argo's wrappers described in \autoref{sec:json-family}.  

\textbf{RQ5 protocol:} In RQ5, we manually investigate clients for which we obtain test failures after library substitution. We re\-run failing tests with a debugger to investigate the root cause of their failure.
From these analyses, we draw principles that enable developers to decouple clients from their libraries. We discuss these principles and give examples of tests that do not respect them.

\section{Results}
\label{sec:results}

%\nicolas{For each RQ answer, start by stating the goal to evaluate in the context of generic Library Substitution Framework, and then explain how we do it on particular on argo}

\subsection{\RQone}
\label{sec:argo-design}

%intro
%In this section, we describe the design process we followed to implement \argo, a Library Substitution Framework for \json libraries.
%We focus on 4 / 20 lib
%We focus on feature that parse (represent) and serialize JSON
%We now detail the process of implementation for bridge, the facade, and the wrappers.

In this research question, we assess the feasibility of implementing a Library Substitution Framework that follows the architecture described in \autoref{sec:archi}. To evaluate the development cost of such an endeavor, we implement a concrete framework, \argo, for the reservoir of \json libraries described in \autoref{sec:json-family}. We focus on the features that are shared by all the $20$ libraries of the reservoir: parsing \json text into Java objects and serializing these objects back to \json text. In the remaining of this section, we discuss in detail our design decisions. We also discuss the effect of our proposed architecture on the development effort through the lens of API elements adapted. 

%\autoref{tab:libs-rq1} 
\begin{table}[ht]
\centering
\rowcolors{2}{gray!25}{white}
\begin{tabular}{lrrrrrr}
  %\toprule
\textsc{Library} & \rotatebox{45}{\textsc{\#API\_Elements}} & \rotatebox{45}{\textsc{\#API\_Bridge}} & \rotatebox{45}{\textsc{\#API\_Wrapper}} \\ 
  \midrule
gson & 468 & 60 & 26  \\
jackson-databind & 5807 & 79 & 36  \\
json & 301 & 123 & 9  \\
json-simple & 102 & 74 & 8  \\
  \midrule
cookjson & 279 & - & 23  \\
corn & 146 & - & 21  \\
fastjson & 1860 & - & 12  \\
flexjson & 326 & - & 6  \\
genson & 747 & - & 4  \\
jjson & 42 & - & 19  \\
johnzon & 199 & - & 13  \\
json-argo & 221 & - & 34  \\
json-io & 230 & - & 6  \\
json-lib & 651 & - & 5\\
json-util & 296 & - & 2 \\
jsonij & 714 & - & 11  \\
jsonp & 392 & - & 14  \\
mjson & 77 & - & 71  \\
progbase & 312 & - & 2  \\
sojo & 788 & - & 4  \\
   \bottomrule 
\end{tabular}
    \vspace{0.5em}
    \caption{Description of the APIs of the \json reservoir}
    \label{tab:libs-rq1}
\end{table}

\autoref{tab:libs-rq1} summarizes the design decisions taken concerning API curating, in our implementation of \argo.
Column \textsc{\#Version} indicates the number of versions listed by \textit{\url{mvnrepository.com}} for each of the $20$ libraries of the reservoir. We decide to focus on only the most popular version of each of these libraries. We estimate the popularity of library version based on the clients we mine from GitHub in \autoref{sec:clients}.
Column \textsc{\#API\_Elements} gives the number of public elements (constructors, methods and fields), in the most popular version of each library. Each of these elements may be called by any clients of the library.
Column \textsc{\#API\_Bridge} shows the number of API elements adapted in the bridges. These elements' code is rewritten using calls to elements of \argo's facade.
These numbers are provided for the subset of the most popular libraries for which  we have chosen to implement a bridge.
Column \textsc{\#API\_Wrapper} indicates the number of API elements of the original libraries used in \argo's wrappers.
In the remaining of this section, we discuss these numbers with more details by going through each layer of the architecture: bridges, facade and wrappers.

%\textbf{Bridges:} As described in \autoref{sec:archi}, we do not need to support all $20$ libraries to support most clients. We can focus on the most popular ones. 
\textbf{Bridges:} We select \nbsrcjson out of the $5$ most popular  libraries in the reservoir: \jackson, \gson, \orgjson, and \jsonsimple. Bridges are implemented by replacing calls to parsing and serialization to their abstract counterparts described in the facade, and by replacing \json data structure classes by classes containing a type from the facade and redirecting method calls to their equivalent described by the facade on the contained type.
We implement a bridge from the most popular version of those \nbsrcjson libraries towards our facade. The column \textsc{\#API\_Bridge} of \autoref{tab:libs-rq1} shows the number of API elements adapted for each bridge. They range from $60$ for \gson to $123$ for \orgjson. This represents a small proportion of their respective API size (ranging from $102$ elements up to \np{5807}). The sources of our bridges are available online.\footnote{\url{https://github.com/nharrand/argo}}

\textbf{Facade:} We identify a common set of features for all of our \nbsrcjson bridged libraries. We rely both on knowledge of the \json subject, and on the manual analysis of the libraries in the reservoir to identify three key, core features, which we specify in  the facade: parsing \json text into Java structures,  serializing these structures to \json text, and representing \json types with Java structures. 
We observe that $15$ out of $20$ libraries propose a type for \json object and $14$ for \json arrays~\cite{harrandjson}. The other libraries represent these data as types of the Java standard library, implementing either a Map or a List (or in one case a primitive array). Therefore, we include  these types in the facade.
We also add a type to represent the \json value Null.%, as some library use the Java value \texttt{null} to represent it, but some others use this value to represent missing values. In order to make these libraries inter-operable it is important to differentiate both of this usage and wrap the \json value Null as a specific value that can not be confused with something else. 
As the other \json types (String, Number, and Boolean) are more often represented by types from the Java standard library, we do not impose any choice in our facade. In total, the facade includes $3$ interfaces (\json object, \json array and \json factory), regrouping $15$ methods to be implemented by each of the $20$ wrappers.

\textbf{Wrapper:} Writing a wrapper consists in implementing the interfaces provided by the facade with the API of an existing library of the reservoir. The column \textsc{\#API\_Wrapper} of \autoref{tab:libs-rq1} shows the number of API elements of the targeted library used in our wrapper. They range from $2$ for \texttt{progbase} and \texttt{json-util} up to $48$ for \texttt{mjson}.
We first implement the interface, described in the facade, responsible for \json text parsing and the creation of concrete \json object from the library being wrapped. This operation is either implemented through the constructors of the class representing \json object and \json array, in a factory class or in a class representing the parser. Second, we adapt serialization operation to serialization methods of the facade. Finally, we add an implementation for the interface of the facade representing \json object and \json array based on the existing class of the library, either by extending the class in question or by creating a container type. When no such class exists, we create a container type for the class of the Java standard library used by the library. This is the case for $5$ libraries: \texttt{flexjson}, \texttt{genson}, \texttt{json-util}, \texttt{progbase} and \texttt{sojo}, which explains the relatively low number of API elements used. In the most extreme case, \texttt{progbase}, only two API elements are used: \texttt{JSONObjectReader.readJSON}, the parser and \texttt{JSONObjectWriter.writeJSON}, the serializer. We also intercept \texttt{null} values used to represent the \json Null value and replace them by the type of the facade.
We do write this wrapper for all $20$ libraries in our dataset. The sources of our wrappers are available online.\footnote{\url{https://github.com/nharrand/argo}}

With the \json reservoir presented in \autoref{tab:libs-rq1} a naive approach for the development of adapters would require the development of an adapter from each API element of each version of the $20$ libraries towards the $19$ other libraries. 
Assuming that the number of API elements for each library does not drastically vary among versions, then, the naive approach would require $27$ millions of elements to be adapted following the calculus (1) presented in \autoref{sec:archi}. Even if we focused on one version per library, this would still mean adapting $13,958$ API elements to each of the $19$ other libraries, or a total of $265,202$ adapters.
Instead, using the revised calculus (2), we adapt a total of $336$ elements in our $4$ bridges, and $20 \times 15$ ($300$) in our wrappers. This represents a total of $636$ adapters. 
This reduction from $265,202$ to $636$ adapters results from a careful selection of a subset of libraries and APIs to build the bridges of \argo, as well as from the architecture based on inversion of control and an abstract facade. The experiments in the following sections show that the drastic reduction of API elements still allows diversifying the supply chain of a large number of clients.

\begin{mdframed}[style=mpdframe]\textbf{Answer to RQ1.}
With \argo, a concrete Library Substitution Framework for the reservoir of \json libraries, we demonstrate that such an implementation is feasible. After API curating, we design bridges that adapt the most used API elements representing the common features of the reservoir. This process reduces the number of API element adaptations from $265,202$ to $636$. %The following research questions assess the impact of this compromise.
\end{mdframed}

\subsection{\RQtwo}

%The goal of a Library Substitution Framework is to generate a maximum of variants from applications that depend on a library from a specific reservoir. These variants must  display an equivalent, yet diverse, behavior from the original application. Hence, the behavior of the artifacts replacing a library (bridge, wrapper and the alternative library), needs to provide diverse yet equivalent behavior. To assess the equivalence of behavior, we propose to take advantage of the existing test suite of the bridged libraries. We run these test suite through the corresponding bridge and wrappers on each  alternative library in the reservoir.
%In this research question, we run the test suite of the original library bridged in \argo through the corresponding bridge and wrappers on each alternative library in the reservoir.

%In this work, we focus on \argo, the Library Substitution Framework described in the previous research question. 
We investigate how \argo's $20$ libraries preserve an equivalent behavior, modulo test, for the $4$ bridges we developed. 
We first discuss the outcome of curating the original test suites of the $4$ bridged libraries.
Second, we run the curated test suites with the bridges and each library of the reservoir, and discuss the outcome.

\subsubsection{Test Selection}

\begin{table}[ht]
\centering
\begin{tabular}{@{}lrrrr@{}}
  \textsc{Library} & \rotatebox{45}{\textsc{\#ORIGINAL}} & \rotatebox{45}{\textsc{\#REMOVED}} & \rotatebox{45}{\textsc{\#PLACEBO}} & \rotatebox{45}{\textsc{\#SELECTED}} \\ 
  \midrule
  gson & 1,051 & 0 & 918 & 133\\ 
  jackson & 2,473 & 0 & 2,332 & 141\\ 
  json & 328 & 230 & 9 & 89\\ 
  json-simple & 29 & 1 & 1 & 27\\ 
  \midrule
  Total & 3,881 & 231 & 3,260 & 390\\ 

  \bottomrule
\end{tabular}
\vspace{1em}
\caption{Results of test selection to assess the correctness of \argo.}
\label{tab:test-selection}
\end{table}

%Original
%intention
\autoref{tab:test-selection} gives the details of the test selection for each original test suite. The second column gives the number of tests present in that original test suite. The third column gives the number of tests that were removed because they call API elements not supported by the bridge. The fourth column indicates the number of tests that we discard because they pass on the \placebo wrapper and therefore would pass with any library. The last column indicates the number of tests that we have selected to run with each library in the reservoir.
For example, the \gson test suite originally contained \np{1051} tests, from which we removed none for compilation reasons, $918$ were labelled as \textsc{PLACEBO} and $133$ were selected.

In total, we assembled $4$ curated test suites combining $390$ tests. %\benoit{we filter out 90\% of the tests, which looks a lot to assess behavior preservation; it would be good to put this number in perspective, e.g., compared to the number of API members in the bridge}
We can see two different trends. \gson and \jackson are large libraries that expose many features, beyond just \json serialization, parsing and \json data structure.  Some of these features are unmodified in our bridges as they are not shared by the other libraries, and therefore unaffected by \argo. By unmodified, we mean that we keep the source code from the original library. This explains the large number of \textsc{PLACEBO} tests that we discard. Note that, by definition, those tests would pass with any wrapper of \argo. On the other hand, \orgjson proposes a large API that we do not fully support in our bridge as few clients rely on it, hence the large number of tests that do not compile because of features that we do not support. Those tests do call API elements that we dismissed in our bridges.

%\benoit{given the wide scope of the paper, we need to discuss the necessity of this deep dive into the test suite; if we keep this, we need to frame it wrt the rest of the paper: test modifications are about behavior specification clarification; how does test removal refer to the concept of lib subst. framework?}
%In the following paragraphs, we give examples of test cases that we have modified or removed because they assess some behaviors that we cannot support in \argo.

%The following paragraph discusses an example of a modified test from our curated test suite. The test assesses a behavior that we cannot support in \argo. It illustrates to what extent behaviors of the libraries can differ on limited aspects.

\begin{lstlisting}[language=java, caption={Test excerpt from \orgjson checking an error message, modified in our curated test suite.}, label={lst:error-handling}]
JSONParser parser = new JSONParser();
s="{\"name\":";
try {
  obj = parser.parse(s);
} catch(ParseException pe) {
  //ORIGINAL
  assertEquals(
    ParseException.ERROR_UNEXPECTED_TOKEN,
    pe.getErrorType());
  //ERROR_HANDLING
  //assertEquals(8, pe.getPosition());
}
\end{lstlisting}

We modify $17$ tests, across the $4$ test suites,
by deactivating assertions that specify the error behavior of the original library. Error handling behavior is very diverse among our libraries and, hence, difficult to reproduce with every other library.
\autoref{lst:error-handling} is an excerpt of a test that checks the behavior of \jsonsimple's parser when encountering an ill-formed \json input. When \jsonsimple's parser encounters an error, gives its position to its client. The client can then decide what to do with the partially parsed \json text. Some libraries, such as \orgjson, fail without giving the position of the error. Other libraries, consider the error to be positioned somewhere else. Hence, it is possible to preserve the fact that an exception is thrown for an ill-formed input, but not necessarily the error message or partially parsed data. Therefore, we keep the assertion that specifies that an exception is thrown (Line 7), but we deactivate the assertion that specifies the content of the exception (Line 11).

%\input{listings/Ill-formed}

%\autoref{lst:forgiving} is extracted from a test that enforces the forgiving nature of the \json parser (in the test suite of \jsonsimple). The string \texttt{s} contains an extra comma, which does not follow the \json grammar specified in RFC 8259~\cite{rfc8259}. Yet, the developers of \jsonsimple have decided to tolerate such ill-formed inputs. Meanwhile, as shown by \textit{Harrand et al.}~\cite{harr2019analyzing}, there exists many ways to handle ill-formed cases among \json libraries. Therefore, we deactivate such assertions as \argo cannot guarantee a common behavior.

\subsubsection{Behavior Preservation}

\autoref{tab:cross-test-results} presents the outcomes for the curated test suites  on each pair of bridge and wrapper in \argo.
Each column corresponds to one of the \nbsrcjson bridges, and each line corresponds to one of the \nbtargetjson wrappers. Each cell gives the number of tests of the original test suite that pass with the bridged library, indicating behavior preservation. Cells in green show wrappers that passed all tests, cells colored in yellow represent wrappers that failed less than $10\%$ of the test suite, and in red are the wrappers that failed more than $10\%$ of the test suite. For example, the first line shows that all the curated tests for \gson pass with the \gson bridge and the \texttt{cookjson} wrapper; $6$ out of the $141$ curated \jackson tests fail with the \jackson bridge and the \texttt{cookjson} wrapper;  all the  \texttt{json} tests and the  \jsonsimple tests pass with their respective bridges and the \texttt{cookjson} wrapper.

\begin{table}[ht]
\centering
\begin{tabular}{@{}lrrrr}
  \hline
\textsc{Implementation} & \textsc{gson} & \textsc{jackson} & \textsc{json} & \textsc{json-simple} \\ 
  \hline
  cookjson & \cellcolor{green!25}133 / 133 & \cellcolor{yellow!25}135 / 141 & \cellcolor{green!25}89 / 89 & \cellcolor{green!25}27 / 27 \\ 
  corn & \cellcolor{green!25}133 / 133 & \cellcolor{yellow!25}136 / 141 & \cellcolor{yellow!25}82 / 89 & \cellcolor{green!25}27 / 27 \\ 
  fastjson & \cellcolor{green!25}133 / 133 & \cellcolor{green!25}141 / 141 & \cellcolor{green!25}89 / 89 & \cellcolor{green!25}27 / 27 \\ 
  flexjson & \cellcolor{green!25}133 / 133 & \cellcolor{green!25}141 / 141 & \cellcolor{yellow!25}88 / 89 & \cellcolor{green!25}27 / 27 \\ 
  genson & \cellcolor{green!25}133 / 133 & \cellcolor{green!25}141 / 141 & \cellcolor{green!25}89 / 89 & \cellcolor{green!25}27 / 27 \\ 
  gson & \cellcolor{green!25}133 / 133 & \cellcolor{yellow!25}137 / 141 & \cellcolor{green!25}89 / 89 & \cellcolor{green!25}27 / 27 \\ 
  jackson-databind & \cellcolor{green!25}133 / 133 & \cellcolor{green!25}141 / 141 & \cellcolor{yellow!25}86 / 89 & \cellcolor{green!25}27 / 27 \\ 
  jjson & \cellcolor{yellow!25}131 / 133 & \cellcolor{yellow!25}131 / 141 & \cellcolor{red!25}71 / 89 & \cellcolor{yellow!25}25 / 27 \\ 
  johnzon & \cellcolor{green!25}133 / 133 & \cellcolor{green!25}141 / 141 & \cellcolor{green!25}89 / 89 & \cellcolor{green!25}27 / 27 \\ 
  json & \cellcolor{yellow!25}132 / 133 & \cellcolor{green!25}141 / 141 & \cellcolor{green!25}89 / 89 & \cellcolor{green!25}27 / 27 \\ 
  json-argo & \cellcolor{green!25}133 / 133 & \cellcolor{green!25}141 / 141 & \cellcolor{green!25}89 / 89 & \cellcolor{green!25}27 / 27 \\ 
  json-io & \cellcolor{green!25}133 / 133 & \cellcolor{green!25}141 / 141 & \cellcolor{yellow!25}88 / 89 & \cellcolor{green!25}27 / 27 \\ 
  json-lib & \cellcolor{green!25}133 / 133 & \cellcolor{green!25}141 / 141 & \cellcolor{green!25}89 / 89 & \cellcolor{green!25}27 / 27 \\ 
  json-simple & \cellcolor{green!25}133 / 133 & \cellcolor{green!25}141 / 141 & \cellcolor{yellow!25}88 / 89 & \cellcolor{green!25}27 / 27 \\ 
  jsonij & \cellcolor{green!25}133 / 133 & \cellcolor{green!25}141 / 141 & \cellcolor{green!25}89 / 89 & \cellcolor{green!25}27 / 27 \\ 
  jsonp & \cellcolor{yellow!25}132 / 133 & \cellcolor{yellow!25}134 / 141 & \cellcolor{green!25}89 / 89 & \cellcolor{green!25}27 / 27 \\ 
  jsonutil & \cellcolor{green!25}133 / 133 & \cellcolor{green!25}141 / 141 & \cellcolor{yellow!25}88 / 89 & \cellcolor{yellow!25}26 / 27 \\ 
  mjson & \cellcolor{red!25}113 / 133 & \cellcolor{red!25}119 / 141 & \cellcolor{red!25}61 / 89 & \cellcolor{yellow!25}25 / 27 \\ 
  progbase-json & \cellcolor{green!25}133 / 133 & \cellcolor{green!25}141 / 141 & \cellcolor{red!25}79 / 89 & \cellcolor{green!25}27 / 27 \\ 
  sojo & \cellcolor{green!25}133 / 133 & \cellcolor{green!25}141 / 141 & \cellcolor{yellow!25}87 / 89 & \cellcolor{yellow!25}25 / 27 \\ 
  \hline
  Total successes & 16 / 20 & 14 / 20 & 10 / 20 & 16 / 20 \\ 
   \hline
\end{tabular}
\vspace{0.5em}
\caption{Behavior preservation matrix for the \nbsrcjson bridges on the \nbtargetjson implementations  in \argo}
\label{tab:cross-test-results}
\end{table}

Overall, out of the $80$ test suite runs ($4 \times 20$), $56$ pass without any failures. 
This means that, $56$ bridge/wrapper couples preserve the initial library's behavior. This shows that for the most part, the libraries of our reservoir share similar features with similar behaviors, which makes them prone to substitution.
$19$ test suite fail less than $10\%$ of their tests, including $7$ that fail a single test. $5$ test suites fail more than $10\%$ of their tests.
Two wrappers (\textit{mjson} and \textit{jjson}) fail tests for all $4$ test suites. While $4$ wrappers (\textit{fastjson}, \textit{genson}, \textit{json-lib}, and \textit{jsonij}) pass all tests of all test suites. These results show that overall, most features of the original four bridged libraries are supported by \argo.

Our analysis of the test failures reveals that libraries can fail for several reasons. Firs, a standard may leave room for diversity among libraries implementing it~\cite{harrandjson}. In the case of \json, the RFC~\cite{rfc8259} leaves \json parsers free to decide whether to accept ill-formed \json inputs. Therefore, tests specifying the acceptance or rejection\footnote{\url{https://github.com/google/gson/blob/f649e051411e092f0123878e16c5132500a2d01e/gson/src/test/java/com/google/gson/internal/bind/JsonTreeWriterTest.java\#L131}} of ill-formed inputs may fail when running on other libraries implementing different choices.
Second, libraries may implement different, and not fully aligned, standards. For example, \texttt{jsonutil} implements ECMA5 and rejects keys that start with a digit\footnote{\url{https://github.com/billdavidson/JSONUtil/blob/f833a1bd7e63df30aae70984539378fdb0ee4263/JSONUtil/src/main/java/org/kopitubruk/util/json/JSONParser.java\#L119}}.
Finally, some aspects of a standard may be commonly ignored by libraries (in our case, $6$ libraries do not support Unicode characters).

\begin{mdframed}[style=mpdframe]\textbf{Answer to RQ2.}
For $56$ out of the $80$ \argo pairs of bridges and wrappers, all test cases in the curated test suite of the original bridged library pass. Moreover, less than 10\%  of the tests fail for $19$ pairs. This is strong evidence that \argo preserves behavior through library implementation substitution and that the developers of future Library Substitution Frameworks can reuse existing test suites for validation.
\end{mdframed}

\subsection{\RQthree}
%79 / 5807 jackson
%60 468 gson
%123 301 json
% 74 102 json-simple
%\benoit{in this first paragraph, and in the 'box' at the end, it's good to remind the reader about the ratio of API elements that we include in the bridges}

%A key aspect of the architecture for Library Substitution Frameworks that we propose, is the  trade\-off  between the subset of the original APIs for which to provide adapters and the share of clients of these APIs for which the library can be substituted. 
%We study the consequences of this trade\-off in the case of our concrete framework, \argo. 
In RQ1, we have discussed the API curating process. In \jackson's bridge we have adapted $79$ out of \numprint{5807} original API elements, for \gson $60$ out of $468$, for \orgjson $123$ out of $301$ and for \jsonsimple $74$ out of $102$. In this research question, we investigate the impact of this API curation on clients of those APIs. In particular, we assess what share of the clients of the $4$ bridged libraries only use API elements supported in \argo. To assess this share, we observe the outcome of compiling the source code of clients of the datasets described in \autoref{sec:clients}, when their \json library is substituted by the corresponding bridge.

%A key design choice for \argo is to decide on a trade\-off  between the subset of the original JSON libraries' APIs present in \argo and the number of client applications for which we can diversify the JSON providers. The goal of \argo is not to support $100\%$ of \json libraries' clients, but to focus on the majority of clients that use the same part of the API. 

%In \jackson's bridge we have adapted $79$ out of \numprint{5807} original API elements, for \gson $60$ out of $468$, for \orgjson $123$ out of $301$ and for \jsonsimple $74$ out of $102$. In this research question, we study the consequences of this trade\-off. We replace the dependency towards the original JSON library by the corresponding bridge. Then, we build the client. A successful compilation indicates  that the JSON API members used by the clients are present in the bridge.

%\input{tables/compilation}

\begin{figure}[ht]
    \centering
    \definecolor{gson}{RGB}{248,118,109}
\definecolor{jsonSimple}{RGB}{199,123,255}
\definecolor{jackson}{RGB}{124,174,0}
\definecolor{json}{RGB}{4,191,196}

\begin{tikzpicture} 
\begin{axis}[xbar,
ymin=0.5,
ymax=4.5,
xmin=0,
xmax=100,
width=0.4\textwidth,
height=4.5cm,
bar shift=0pt,
bar width=4.5mm,
axis x line* = bottom,
axis y line* = left,
xtick style={draw=none}, % Hide tick line
ytick style={draw=none}, % Hide tick line
label style={font=\small},
tick label style={font=\small},
visualization depends on=x \as \XVal,
nodes near coords={%
\small\pgfmathprintnumber\XVal\%
},
every node near coord/.style={
  align=center,
  color = black,
  anchor=east
},
yticklabels={json-simple,json,jackson-databind,gson},
ytick={1,2,3,4},
xmajorgrids = true,
xlabel=Share of clients that successfully compile (\%),
]

\addplot+[jsonSimple] coordinates {(85.29,1)};
\addplot+[json] coordinates {(80.95,2)};
\addplot+[jackson] coordinates {(89.84,3)};
\addplot+[gson] coordinates {(95.90,4)};

\end{axis}
\end{tikzpicture}

% json-simple &  34 &  29 (85.29\%) \\ 
% json &  84  &  68 (80.95\%)\\ 
% jackson-databind & 128 & 115 (89.84\%) \\
% gson & 122 & 117 (95.90\%)  \\ 
    \caption{Share of client projects with no compilation errors after transformation}
    \label{fig:compilation}
\end{figure}
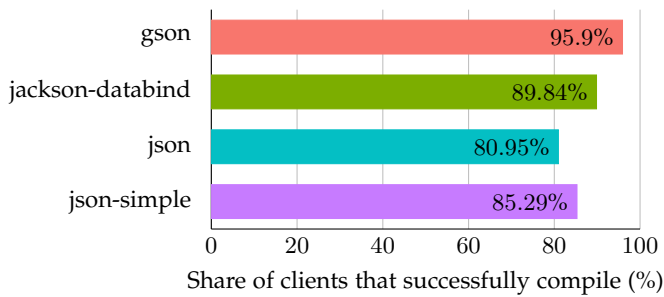

\autoref{fig:compilation} shows the compilation outcome of $368$ clients transformed with one of the \nbsrcjson bridges in \argo. The horizontal axis indicates the bridge, and the vertical axis the share of clients using the bridge that compile without error.
%\autoref{tab:compilation} shows the compilation outcome of $368$ clients transformed with one of the \nbsrcjson bridges in \argo.
%The column `\#Compile' indicates the number of clients that successfully compile with the bridge, the number in brackets is the ratio of successful compilations for a bridge. 
For example, $115$ of the $128$ ($89.84\%$) clients that originally use \jackson compile with \argo's bridge for \jackson (\texttt{jackson-databind-over-argo}).

In total, $89.40\%$ of clients ($329$ out of $368$) successfully compile with an \argo bridge instead of their original \json dependency. This is strong evidence that the trade-off that we made between the complexity of the bridges and the preservation of a subset of the original APIs is appropriate to successfully serve a large portion of the JSON libraries clients. 
In the following, we discuss the root cause for the compilation errors. In particular, we highlight that not all of them are due to missing API members in the bridges. Instead, they occur because of the intrinsic complexity of dependency resolution mechanisms.

%Incomplete API
%\emph{Targeted bridge API}.
By design, \argo bridges do not cover the most exotic parts of the bridged libraries' API. Hence, the compilation fails when we build a client that uses an API element that is not supported in the bridge. This is the case, for example, for Jooby\footnote{\url{https://github.com/jooby-project/jooby}}, a modular web framework that depends on \orgjson. Jooby depends on the class \texttt{org.json.JSONTokener}, which is not included in our bridge. Hence, when replacing \orgjson, by \texttt{json-over-argo}, compilation fails.
Among the $16$ \orgjson clients for which \argo transformations lead to compilation errors, $10$ are due to clients relying on \texttt{org.json.JSONTokener}.

%Old API
%\emph{\json library version}.
We provide a bridge for the most used version of each bridged library. Consequently, clients relying on a different version of the library may face some incompatibility when the API is substituted by the bridge. This only causes issues when the part of the API used by the client has changed in the version supported by \argo. For example, the project java2typescript\footnote{\url{https://github.com/raphaeljolivet/java2typescript/blob/5e3921b674bfb03ba9ac042238a8db0b9b17ca3a/java2typescript-jackson/pom.xml}} depends on \jackson version $2.6.4$. It uses the constructor for the class \texttt{TypeBindings}, available in version $2.6.4$. We support \jackson version $2.12.0$ in which this constructor has become private, providing a factory method instead to create an object of this type. Hence, when replacing \jackson by its bridge \texttt{jackson-databind-over-argo} in java2typescript, the compilation fails.

%Dependency conflict
%\emph{Dependency conflict}.
Compilation errors occur because of dependency conflicts introduced by specific implementation choices of \argo. For example, \texttt{netbeans-mmd-plugin} depends on \texttt{commons-codec:1.14}. But \textit{cookjson}, one of the libraries in our JSON reservoir, depends on version \texttt{1.10} of \texttt{commons-codec}. This creates a conflict in the compilation classpath of the project when \orgjson is substituted by its \argo bridge and the \textit{cookjson} implementation. This conflict triggers a compilation error, since the API of \texttt{commons-codec} has changed between version \texttt{1.14} and \texttt{1.10}. In particular, the method \texttt{DigestUtils.digest}\footnote{\url{https://github.com/apache/commons-codec/blob/af7b94750e2178b8437d9812b28e36ac87a455f2/src/main/java/org/apache/commons/codec/digest/DigestUtils.java\#L71}}'s signature has changed and since version \texttt{1.11} takes  a \texttt{byte[]} parameter instead of an \texttt{InputStream}. This leads the compilation of \texttt{netbeans-mmd-plugin}, that relies on the newest version of the API, to fail.

%79 / 5807 jackson
%60 468 gson
%123 301 json
% 74 102 json-simple

As stated in \autoref{sec:argo-design}, the goal of our architecture is not to target all clients of all versions of the libraries in our reservoir. We have focused on a limited part of the API of one version of $4$ libraries. We observe that we still provide adaptation for API elements that are enough to support $89.40\%$ of the clients of these $4$ libraries.

\begin{mdframed}[style=mpdframe]\textbf{Answer to RQ3.} The \nbsrcjson \json bridges implemented in \argo provide an API large enough to successfully compile $89.40\%$ of the clients in our dataset. 
%Compilation errors can happen when clients use an unusual part of the original API or when they depend on a different version of the library supported by the Library Substitution Framework. 
The high rate of successful compilation is evidence that it is relevant to select a subset of a library's API to build a Library Substitution Framework based on bridges and wrappers. 

\end{mdframed}

\subsection{\RQfour}

In the section, we evaluate the amount of behavior preserving variants per client, that we can generate through the use of a Library Substitution Framework. To assess which variant exhibit a behavior equivalent to the original application, we rely on the test suite of the application.

%We carry this investigation on the concrete framework we have implemented in RQ1, \argo. For this experiment 
We use the $329$ clients of the dataset described in \autoref{sec:clients} for which the substitution of the \json library by its corresponding bridge led to no errors. For each of these clients, we build $20$ variants corresponding to each of the $20$ libraries of the reservoir and run the client's test suite. We then analyze the test results and count the number of variant that are behaviorally equivalent modulo test.
We ran the test suite of each client three times to limit the impact of flaky tests, the clients have \np{960074} test cases, which take a total of \np{18.2} hours to execute. In total, we execute \np{2880222} tests cases.  

%In this section, we investigate how many alternative \json libraries \argo can substitute in clients. We build each client with each of our \nbtargetjson \json library wrappers can be successfully included in each client's build. We consider that an alternative \json library can be successfully substituted in the client if all of its test cases pass. %\david{How many experiments is that and how long time did it take to both compile and execute the test suits of all combinations?}

\begin{figure}[ht]
    \centering
    \includegraphics[width=\columnwidth]{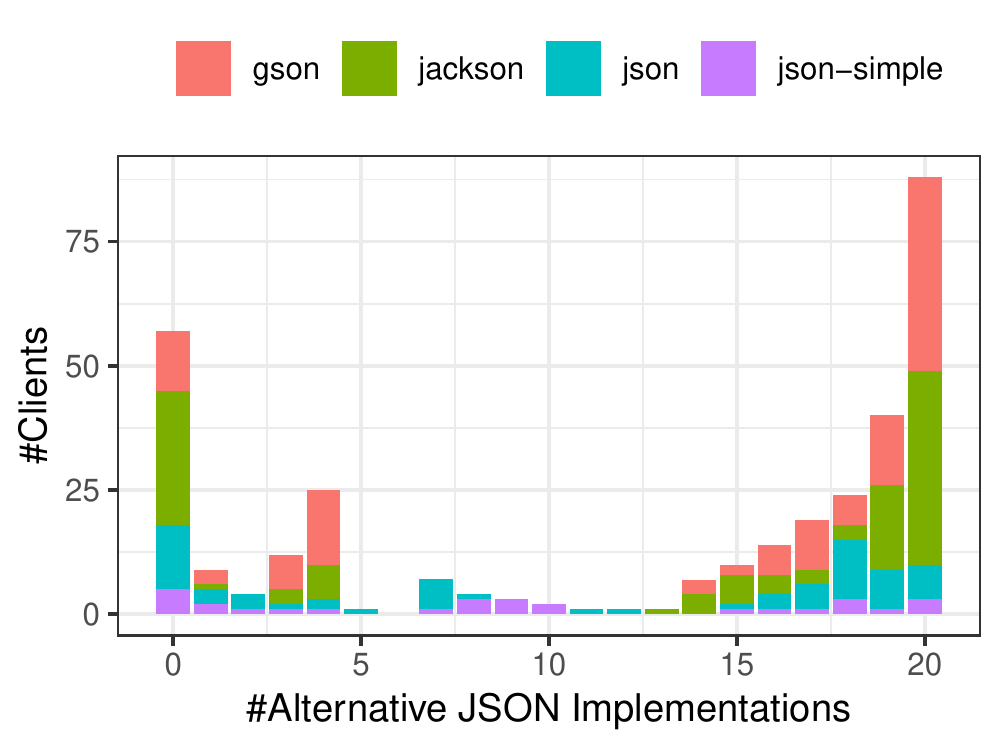}
    \caption{Distribution of the number of successful substitution per client.\\ We regroup the $329$ clients of our dataset based on the number of \json library substitutions (out of $20$) that passed their test suite.}
    \label{fig:number-alternative}
\end{figure}
%\david{For Fig 8., I am wondering if it is easier to look at a graph that number of \todo{implementations} that at least are successful. Maybe, maybe not.}

\autoref{fig:number-alternative} illustrates the key results for RQ3. 
This plot shows the cumulative number of clients for which we successfully substituted a certain number  of alternative \json libraries. The horizontal axis gives the number of alternative libraries and the vertical axis gives the number of clients. The colors indicate which bridge is used.

The right-most bar of the plot indicates that \argo is able to substitute the original \json library by any of the $20$ alternatives for $88$ clients (including the wrapper corresponding to the original library). 
This is a novel evidence that it is possible to automatically increase the diversity of suppliers in the supply chain of client applications, without changing the code of these clients. Among the $88$ clients, $39$ build with the \gson bridge and $39$ build with the \jackson bridge, $7$ with the \orgjson bridge and $3$ for the \jsonsimple one. This shows that library substitution are possible from any of the $4$ bridged libraries towards any of the $20$ libraries of the reservoir.

The other bars in the plot indicate that the majority of the other $241$ clients can be provided with a large proportion of alternative \json libraries available in our pool: $195$ out of $329$ ($59.27\%$) clients can build with $14$ or more alternatives.
On the left of the plot, we observe that $57$ clients cannot successfully run their whole test suite on any alternative library ($12$ for \gson, $27$ for \jackson, $13$ for \json and $5$ for \jsonsimple), and that 
$51$ clients are compatible with between $1$ and $5$ \json libraries.

In total, we successfully synthesize \np{4069} behaviorally correct variants, modulo  test suite, for the  the $329$ original clients. We can build variants, with $15$ or more alternative \json libraries, for \nbmassdivclient clients. In the following, we discuss two clients for which a high number of libraries can be substituted to their original \json library.

The Automotion framework\footnote{\url{https://github.com/ITArray/automotion-java}} originally depends on \jsonsimple. Its test suite includes $5,693$ test cases that trigger invocations on the  \jsonsimple API. Automotion directly calls $15$ different methods from $3$ classes of the library (\texttt{JSONObject}, \texttt{JSONArray} and \texttt{JSONValue}.  These calls are made from $13$ distinct methods in \texttt{automation\-java}'s code. \json manipulation is important for this framework (evidenced by the large number of test cases), yet it is very regular: it uses functions related to \json parsing, \json object manipulation (read and write) and serialization. During the execution of its test suite,  Automotion invokes \jsonsimple's API   $39,481$ times. All $20$ libraries provided by \argo can be substituted to \jsonsimple and provide the proper behavior for these thousands of invocations, without breaking the test suite of Automotion.

\begin{lstlisting}[caption={Method from \texttt{corn} detecting the correct Java type to \\ represent a \json literal}, label={lst:corn-detect-class}]
private static Class<?> 
    detectClass(String str) {
  try {
    Integer.valueOf(str.trim());
    return Integer.class;
  } catch (Exception ex) {}
  try {
    Float.valueOf(str.trim());
    return Float.class;
  } catch (Exception ex) {}
  try {
    Double.valueOf(str.trim());
    return Double.class;
  } catch (Exception ex) {}
  try {
    new BigInteger(str.trim());
    return BigInteger.class;
  } catch (Exception ex) {}
  try {
    new BigDecimal(str.trim());
    return BigDecimal.class;
  } catch (Exception ex) {}
  return String.class;
}
\end{lstlisting}

CrashCoin\footnote{\url{https://github.com/StanIsAdmin/CrashCoin}}, a cryptocurrency implemented in Java, originally depends on \orgjson. CrashCoin's test suite includes $15$ test cases that cover parts of CrashCoin which invoke the API of \orgjson. For all the libraries in our reservoir, except \texttt{corn} and \texttt{mjson}, we successfully build CrashCoin with \argo and all test cases pass, which generates 18 variants of the project with 18 different providers for \json processing. 
CrashCoin's test suite fail when substituting \orgjson by \texttt{corn}.
\texttt{TestBlockChain\#testBlockChainBlockAdding} fails with \texttt{corn} because this \json  library does not handle correctly natural numbers that do not fit in an \texttt{Integer} (number in the interval $2^{31} - 1$ to $-2^{31}$). 
When running the test, the \json library needs to determine the type of $1,512,901,875,251$. The library \texttt{corn} relies on the method \texttt{detectClass}, shown in \autoref{lst:corn-detect-class}, to determine the  Java type to represent a \json literal. This method tries to parse the literal as \texttt{Integer}, if this fails, it tries to parse the literal as a \texttt{Float}, then as a \texttt{Double}, and so on. But parsing the String $1,512,901,875,251$ with \texttt{Float\#valueOf} does not raise an exception for values that do not fit on 32 bits. Instead, the method returns the rounded value $1.51290184E12$, losing the last $5$ digits of precision and failing the test.

Overall, the $2$ libraries that pass the least clients' test suite, \texttt{mjson} ($120$ out of $329$) and \texttt{jjson} ($135$ out of $329$) are also the two libraries that fail the most tests in our cross testing experiments as shown in \autoref{tab:cross-test-results}. This confirms the intuition given in RQ1, that these two libraries exhibit a behavior that is more different from the $4$ bridged libraries than the other libraries, hence their use as substitution libraries is less likely to succeed.

\begin{mdframed}[style=mpdframe]\textbf{Answer to RQ4.} \argo successfully diversifies the \json supply chain of $272$ of the $329$ ($83\%$) client applications. \nbmassdivclient of them are compatible with $15$ or more libraries. In total, from $329$ clients, we synthesize \np{4069} variant applications using diverse \json suppliers.
This is an empirical assessment of the supply chain diversity that can be achieved by the  Library Substitution Framework proposed in this work.
\end{mdframed}

\subsection{\RQfive}

%\benoit{which of the following points are specific to \json and which ones can generalize to applications that wish to diversify their supply chain, e.g., relate to the general concept of lib subst fram.?}

In this research question, we analyze the root cause of test failures when testing the clients with \argo. 
A key observation is that some test failures occur because of a tight coupling between the client and the \json library. Some coupling is accidental and can be avoided to eventually make the client more prone to supply chain diversification. In this section, we discuss and illustrate three principles that enable more diversification. While the examples are collected from our experiments on \json library substitution, we draw from them generic principles that are not specific to \json libraries.

\begin{lstlisting}[caption={Excerpt of 
CorpusMetadata constructor, from Lambda-3/Indra}, label={lst:cast}]
CorpusMetadata(Map<String,Object> data){
  corpusName    = (String) data.get(CORPUS_NAME);
  applyStemmer  = (long) data.get(APPLY_STEMMER);
  [...]
}
\end{lstlisting}

%  removeAccents = (boolean) data
%        .get(REMOVE_ACCENTS);
%  transformers = 
%    (Map<String,Collection<String>>) data
%    .get(TRANSFORMERS);

\textbf{Minimize reliance on implicit library behavior:} Some undocumented behaviors of a library, that are more linked to its implementation than to the contract of its API, are used by some clients.
\autoref{lst:cast} shows such an example of dependence over an observed behavior rather than a contractual one. \texttt{CorpusMetadata} is class from  Indra\footnote{\url{https://github.com/Lambda-3/Indra}}, which depends on \jsonsimple. The constructor of this class takes a \texttt{Map} as an argument. In practice, this \texttt{Map} is read from a \json object parsed with \jsonsimple. This constructor assumes that specific types correspond to each key. In particular, the entry retrieved for the key \textsc{APPLY\_STEMMER} is assumed to be typed as a \texttt{long}. While this does not cause any issue with numbers read by \jsonsimple, other \json libraries may encode them with a different Java type. For example, \texttt{fastjson} encodes some value with \texttt{Integer}. Hence, when \argo substitutes \jsonsimple with \texttt{fastjson}, the cast operations fail and a \texttt{ClassCastException} is thrown. This illustrates a case of a client that relies on an undocumented behavior of a library (returning a \texttt{Long} for natural numbers) rather than the one guaranteed by the API (returning an \texttt{Object}). Limiting coupling to return types that are not guaranteed by the signature of an API element is a sound design principle that facilitates dependency substitution and diversification.

\begin{lstlisting}[caption={Test excerpt from jts   GeoJsonWriterTest}, label={lst:string-equality}]
private void runTest(String wkt, 
    int srid, boolean encodeCRS, 
    String expectedGeojson) 
    throws ParseException {
  Geometry geom = read(wkt);
  geom.setSRID(srid);
  geoJsonWriter.setEncodeCRS(encodeCRS);
  String json = geoJsonWriter.write(geom);
  json = json.replace('"', '\'');
  assertEquals(json, expectedGeojson);
}
\end{lstlisting}

\textbf{Test cases should not over specify the behavior of the client:} 
\autoref{lst:string-equality} shows an example of a test that over specifies the behavior of the program under test. It is extracted from the test suite of \texttt{jts}\footnote{\url{https://github.com/locationtech/jts}}. It checks that the class \texttt{Geometry} is correctly serialized in \json. However, instead of  checking that the information is indeed contained in a well-formed \json string, the test  checks that the serialized string is exactly equal to an expected value, character per character. This is over specification of the expected behavior, since the \json standard specifies that the order of key-value pairs is not meaningful. In this case, $16$ out of $20$ \argo libraries make this test fail while they still produce an equivalent \json string, which is not strictly equal. The other $4$ libraries pass the test.
Over-specifying tests hinders library substitution, by erroneously indicating to the developer of the client that a library is incompatible.

\begin{lstlisting}[caption={Test excerpt from CouchbaseMock testValueValidation}, label={lst:test-invalid}]
@Test
public void testValueValidation() 
 throws SubdocException {
  assertBadDictValue("INVALID");
  assertBadDictValue("1,2,3,4");
  assertBadDictValue("1,\"k2\":2");
  assertBadDictValue("{ \"foo\" }");
  assertBadDictValue("{ \"foo\": }");
  assertBadDictValue("nul");
  assertBadDictValue("2.0.0");
  assertBadDictValue("2.");
  assertBadDictValue("2.0e");
  assertBadDictValue("2.0e+");
}
\end{lstlisting}

\textbf{Limit dependency on behavior stricter than the standard:} 
%\benoit{how is a non-standard different from an implicit library behavior ? (first point in this section)}
%nicolas: The first point discuss reliance on implementation details. Accepting or rejecting incorrect value is not an implementation detail. It is a deliberate choice by json-simple. Here the client expect a behavior that is more striclty defined than the standard. Hence, other library may both respect the standard, and make this test fail.
\autoref{lst:test-invalid} shows an excerpt of a test for CouchbaseMock\footnote{\url{https://github.com/couchbase/CouchbaseMock}}. It checks that specific values cannot be inserted in a \json object. RFC 8259~\cite{rfc8259}, that standardizes the JSON format, stipulates that JSON libraries should not generate such values when they produce JSON text, but that they are free to accept them or not when reading JSON text. This test from CouchbaseMock verifies that such values are rejected. This means that CouchbaseMock is stricter than the \json standard.
CouchbaseMock depends on \jsonsimple, which rejects these ill-formed values. Yet, other libraries often accept these values~\cite{harrandjson}, without violating RFC 8259~\cite{rfc8259}. Hence, when \jsonsimple is substituted by another \json library, the test may fail. In our dataset, $11$ of the $20$ libraries make this test fail, while the other $9$ make it pass.
We showed in our previous work~\cite{harrandjson} that there exists a large diversity of behavior among \json libraries when parsing ill-formed. This diversity emerges because the behavior of \json libraries is not standardized for ill-formed values. 
It is a sound design principle for client applications to limit their assumptions about nonstandard API behavior, which can favor automatic dependency diversification.

\begin{mdframed}[style=mpdframe]\textbf{Answer to RQ5.} A manual analysis of corner cases, reveals three design principles can favor dependency diversification: (i) Clients should not depend on implicit behavior of the library; (ii) test cases should not over-specify the behavior of the client they test; (iii) clients should limit their assumptions about non-standard behavior.
%A qualitative analysis of test failures among clients for which we substitute the library reveals three design principles that will favor dependency diversification:
\end{mdframed}

%\subsection{\RQfive}

%\todo{rename RQ4 / RQ5}

%We run the json benchmarks with JMH

%\begin{enumerate}
%    \item Argo with bridge and original implementation
%    \item Variations of overhead with Argo with bridge and using other \todo{implementations}
%\end{enumerate}

\section{Threats to validity}
\label{sec:threats}

\textbf{Internal Validity.}
We evaluate the concept of Library Substitution Framework with a concrete implementation for the Java \json reservoir. We rely on application test suites as an oracle for successful substitution. However, test suites are imperfect. 
%In  particular,clients  may  declare  dependencies  towards  a  \json  library that  they  do  not  use  \cite{harr2019analyzing,sotobloat}.  Furthermore,  a  client  may use a library but, the client’s test suite might not cover the library.
To limit the impact of this threat to validity, we rely on a two stages process. \argo is tested with carefully curated test suites of the $4$ bridged libraries. We systematically remove libraries which test suite passes with the \placebo wrapper. This ensures that the tests cover the \json library and that its behavior is actually depended on.

\textbf{External Validity.} We evaluate the concept of Library Substitution Framework in the single context of \json libraries. This opens a discussion on how does this concept apply to other domains.
We foresee three key arguments in favor of such an approach  for other domains.
Reservoirs of similar libraries exist in several domains, as evidenced by  browsing software repositories for libraries of a specific domain\footnote{\url{https://mvnrepository.com/open-source}}. %But this has also been studied extensively in the scientific literature, in particular to recommend alternative libraries to developers~\cite{Sun2020,Bogomolov2020}.
Previous work has demonstrated the possibility of performing wrapper based migration for Swing and SWT libraries~\cite{bartolomei2010swing}, XML~\cite{bartolomei2009study} or  user interfaces~\cite{Bartolomei2012}.
The overall architecture that we propose in this work is inspired by the one of the framework Simple Logging Facade for Java~\cite{slf4j}. Even if the architecture of \texttt{slf4j} does not aim at the same goals, it shows that such an architecture is possible in the domain of logging libraries. %The library \texttt{micrometer}~\cite{micrometer} shows that it applies for metrics as well.

%as mentioned the prerequisite of library reservoir exists for numerous other domains. Secondly \json is not unique in having a standard. 
% finally other work have replaced other thing even if at a different granularity. And slf4j

%\textbf{Construct validity.}

%We both rely on test and extensive manual analysis some of which can be shown in the numerous examples in the papers

%\textbf{Limitations. and future work} Performances
%Features

%\section{Discussion}
%\label{sec:discussion}

%\subsection{Diversify the supply chain beyond JSON}

%\begin{enumerate}
%    \item JSR
%    \item ORM, Collections ?
%    \item crypto APIs
%\end{enumerate}

%\subsection{Design principles to increase diversity in the supply chain}

%\subsection{Diversification in the lifecycle}

%the paper is about build time, but the same techniques can be applied at load or runtime

%cite multi-variant execution, n-version

%\subsection{Supplier diversity and reproducible build}

\section{Related Work}
\label{sec:related-work}
Our work about diversification in the software supply chain relates to three research areas: automatic synthesis of software diversity, library management and software supply chain hardening.

\subsection{Automated Diversity}

Automatic diversification was pioneered by the seminal work of  Cohen~\cite{cohen93} and Forrest \cite{forrest97}. They proposed  to generate program variants that deliver the same functionality as the original, while exhibiting differences in the way they handle unspecified behavior.
Following these initial work, subsequent studies analyzed different kinds of transformations in the stack \cite{lee2021savior},  on binary code \cite{wartell2012binary,AbrathCMBS20}, at the binary interface level \cite{Kc03}, in the compiler \cite{homescu2015large} or in the  source code \cite{baudry14b,harrand2019journey}.
All these transformations operate on a small scope, typically at the instruction-level. This is both a necessity to limit the risks of breaking the original functionality, and a key limitation to achieve behavior diversity outside the specified requirements. 
Meanwhile, only few techniques propose to exploit the existence of alternative libraries to create diversity. Persaud and al.~\cite{persaud2016frankenssl} present FrankenSSL. This  approach  recombines fragments of forks of OpenSSL to create library variants that all provide the same features. But these fork have a very similar API as they descend from a unique project.
Our proposal for a Library Substitution Framework generalizes the idea of exploiting the natural diversity of library implementations, at build time.

The natural emergence of functionally diverse implementations of the same features has been harnessed in several ways in the past, to reduce common failure \cite{SalakoS14}. For example, collection libraries exist in many different implementations, which can be selected according to application specific performance requirements, either statically ~\cite{basios2018darwinian}or dynamically~\cite{shacham2009chameleon,basios2018darwinian}.
We have previously harnessed the natural diversity of Java decompilers to improve the overall precision of decompilation~\cite{HarrandDecompiler}. Gashi and colleagues have tamed the diversity of SQL servers for security purposes \cite{gashi2007}, while Xu and Kim leveraged the execution platform diversity to protect against attacks on PDF applications \cite{xu2017}. 
The concept of Library Substitution Framework is founded on the natural emergence of diverse implementations of various kinds of libraries. Our concrete implementation, \argo, harnesses \json libraries that  differ in performance~\cite{Maeda} as well as in behavior~\cite{harrandjson}. This type of diversity in the supply chain of applications is beneficial to protect against a single point of failure.

\subsection{Library Migration}
%Study actual migrations

Techniques to help developers in the migration process usually suggest mappings between specific API elements providing similar features
Chen et al.~\cite{chen2015} describe an unsupervised learning approach to create a database of similar APIs. 
Alrubye and colleagues~\cite{alrubaye19,Alrubaye18} present a tool named MigrationMiner that finds migration rules, in past migrations of other projects. These rules help developers to replace calls to the previous API by their equivalent in the new API.
Fazzini and colleagues~\cite{Fazzini2020} propose APIMigrator, to automate the migration of  applications when the Android API evolves, based on how other projects have already migrated. 
These tools assist developers in their migration process, and do not automate the migration of applications. 
They could assist the developers of a Library Substitution Framework in the development of bridges.

%Adapters allowing the replacement of an API call by another without client modifications have been proposed before. But none to our knowledge propose to substitute any library of a reservoir by another, for diversification purposes, as in this work.
In his thesis, Bartolomei~\cite{Bartolomei2012} proposes design principles for wrapper-based migration. %This work provides useful architectural patterns that we use to write the bridges implemented in \argo.
Sharma and colleagues~\cite{sharma2019} wrote a tool that synthesizes adapters to replace functions that provide similar features, one by another, inside a given project.
Previous projects \texttt{slf4j}~\cite{slf4j} or \texttt{micrometer}~\cite{micrometer} implement such a family of adapters for reservoirs of logging and metric libraries. 
They aim at merging respectively logs and metric gathered by the different libraries present in the dependency tree of an application.
In this work, we propose a generic architecture for library substitution, which rely on a similar architecture for a different purpose. Our goal here is to build variants of applications, depending on different library suppliers.

%An Analysis of Library Rollbacks: A Case Study of Java Libraries

%From API to NLI: A New Interface for Library Reuse

%Mining Multi-level API Usage Patterns

%Mining Unit Tests for Discovery and Migration of Math APIs ?

\subsection{Supply Chain Hardening}

Recent work has highlighted the need for software supply-chain hardening, with a focus on software dependencies~\cite{PeisertSOMBLMMP21}
%\cite{cox2019surviving,PeisertSOMBLMMP21,MassacciJP21,MassacciP21}. 
One approach consists a systematically building dependencies from source code and ensuring a strictly deterministic build~\cite{lamb2021reproducible}. Deterministic builds are important as discrepancies between sources and packages can hide malicious code from the eyes of reviewers of open-source software~\cite{Vu2021}. But making builds fully deterministic is currently challenging because of timestamping or random naming \cite{ren2018automated}.
Nikitin and colleagues propose a decentralized software-update framework to distribute software packages for which the build is reproducible~\cite{nikitin2017chainiac}.

Other works aim  at defending against malicious libraries also exists. Pashchenko and colleagues~\cite{pashchenko2020vuln4real} propose a methodology to evaluate which dependencies of a software project are vulnerable. 
Vasilakis and colleagues~\cite{vasilakis2021supply} propose a technique that learns the normal behavior of a dependency and automatically synthesizes a new library that exhibits an equivalent client-observable behavior, with no vulnerability. Catuogno and colleagues propose a technique to prevent malicious users from forcing applications to install vulnerable dependencies \cite{CatuognoGP20}.

Cox~\cite{cox2019surviving} suggests that developers of client applications should decouple their code from the concrete implementation of libraries, adding an abstraction layer between the application and the API. This recommended abstraction layer regroups all calls to the external API in a localized wrapper, making future migrations simpler. 
We propose to mutualize this solution to all clients of a library reservoir. Our notions of bridge and facade for a library reservoir detach the APIs of these libraries from their actual implementation, allowing for library substitution.

%Our work

\section{Conclusion}
\label{sec:conclusion}
%\nicolas{to update}

In the context of ever-increasing reliance on external software libraries, supply chain attacks represent a major threat. To mitigate their scale, we propose to generate and deploy software variants with a diverse supply chain.

In this work, we present the concept of Library Substitution Framework based on a three-tier adaptation architecture. It allows developers to substitute a library by an alternative one, without modifications of the client software, enabling diversified variants based on each different library alternative. To assess the validity of the architecture we propose, we implement a concrete framework for Java JSON libraries. Our framework, \argo, supports $20$ different \json libraries. We test this framework by reusing the test suites of bridged libraries.
We evaluate the ability of our framework to diversify the supply chain of applications on a dataset of open-source project depending on \json libraries.
On \nbmassdivclient of the \nballclient java applications tested, we are able to provide at least $15$ alternatives.

These results open the way for usages of the variants produced by our approach such as Moving Target Defense, or N-Variant systems.

%context
%problem
%sol
%argo
%results
%opening

%\cite{harr2019analyzing}

\bibliography{bib}{}
\bibliographystyle{ieeetr}

\end{document}